\documentclass{aa}  

\usepackage{graphicx}
\usepackage{natbib}
\usepackage{graphicx}
\usepackage{txfonts}
\usepackage[hidelinks,colorlinks=true,linkcolor=blue,citecolor=blue]{hyperref} 
\usepackage{mathtools}
\usepackage{subcaption}
\usepackage{balance}
\usepackage{textcomp,gensymb}
\usepackage{amsmath}

\usepackage{txfonts}
\usepackage{comment}
\usepackage{sidecap}

\makeatletter
\renewcommand*\aa@pageof{, page \thepage{} of \pageref*{LastPage}}
\makeatother

\newcommand{\cheops}{\texttt{CHEOPS}\xspace}
\newcommand{\jwst}{\texttt{JWST}\xspace}
\newcommand{\hst}{\texttt{HST}\xspace}
\newcommand{\wfc}{\texttt{WFC3}\xspace}

\newcommand{\nirspec}{\texttt{NIRSpec}\xspace}

\newcommand{\spitzer}{\texttt{Spitzer}\xspace}
\newcommand{\tess}{\texttt{TESS}\xspace}
\newcommand{\plato}{\texttt{PLATO}\xspace}

\newcommand{\Rv}{\ensuremath{R_{\mathrm{v}}}\xspace}

\newcommand{\hf}{\ensuremath{h_{2f}}\xspace}

\begin{document}

\title{On the effects of tidal deformation on planetary phase curves}
\authorrunning{Akinsanmi et al.}

   \author{Babatunde Akinsanmi\
          \inst{1}
          \and
          Monika Lendl\inst{1}
          \and
          Gwenael Boué\inst{2}
          \and
          Susana C.~C. Barros \inst{3,4}
          }

   \institute{Observatoire astronomique de l’Université de Genève, chemin Pegasi 51, 1290 Versoix, Switzerland. 
        \email{tunde.akinsanmi@unige.ch}
   \and
   IMCCE, UMR8028 CNRS, Observatoire de Paris, PSL Univ., Sorbonne Univ., 77 av. Denfert-Rochereau, 75014 Paris, France
   \and
   Instituto  de  Astrofísica  e  Ciências  do  Espaço,  Universidade  do Porto, CAUP, Rua das Estrelas, PT4150-762 Porto, Portugal
   \and Departamento\,de\,Fisica\,e\,Astronomia,\,Faculdade\,de\,Ciencias,\,Universidade\,do\,Porto,\,Rua\,Campo\,Alegre,\,4169-007\,Porto,\,Portugal}
   
   \date{Received 17 August, 2023; accepted 4 October, 2023}

\abstract{With the continuous improvement in the precision of exoplanet observations, it has become feasible to probe for subtle effects that can enable a more comprehensive characterization of exoplanets. A notable example is the tidal deformation of ultra-hot Jupiters by their host stars, whose detection can provide valuable insights into the planetary interior structure. In this work, we extend previous research on modeling deformation in transit light curves by proposing a straightforward approach to account for tidal deformation in phase curve observations. The planetary shape is modeled as a function of the second fluid Love number for radial deformation \hf. For a planet in hydrostatic equilibrium, \hf provides constraints on the interior structure of the planet. We show that the effect of tidal deformation manifests across the full orbit of the planet as its projected area varies with phase, thereby allowing us to better probe the planet's shape in phase curves than in transits. Comparing the effects and detectability of deformation by different space-based instruments (\jwst, \hst, \plato, \cheops, and \tess), we find that the effect of deformation is more prominent in infrared observations where the phase curve amplitude is the largest. A single \jwst phase curve observation of a deformed planet, such as WASP-12\,b, can allow up to 17$\sigma$ measurement of \hf compared to 4$\sigma$ from transit-only observation. Such high precision \hf measurement can constrain the core mass of the planet to within 19\% of the total mass, thus providing unprecedented constraints on the interior structure.
Due to the lower phase curve amplitudes in the optical, the other instruments provide $\leq4\sigma$ precision on \hf depending on the number of phase curves observed. We also find that detecting deformation from infrared phase curves is less affected by uncertainty in limb darkening, unlike detection in transits. Finally, the assumption of sphericity when analyzing the phase curve of deformed planets can lead to biases in several system parameters (radius, day and nightside temperatures, and hotspot offset among others), thereby significantly limiting their accurate characterization.}


   \maketitle
%

\section{Introduction}
In ultra-hot Jupiter systems, the tidal interaction between the star and the planet results in the deformation of both bodies into ellipsoidal shapes. The deformation of the star leads to a periodic modulation of the observed flux as the stellar tide, raised by the planet, rotates with the planet's orbital phase \citep{russel_1912, Morris1985TheStars., mislis2012, Esteves2013OPTICALEXOPLANETS}. This periodic signal in the light curve is referred to as stellar ellipsoidal variation, a phenomenon that has been empirically confirmed in several hot Jupiter systems \citep[see e.g.,][]{shporer-2019-WASP18, wong2021_tess2ndyearPCs, parvieinen2022}. Similarly, the stellar tidal force deforms the planet along the axis pointing toward the star. The tidal deformation of the planet is challenging to measure during transit as the elongated axis is predominantly aligned with the line of sight, thereby resulting in a nearly circular sky projection of the planet. Nonetheless, slight deviations from the standard spherical planet transit light curve can be observed as the shape and cross-sectional area projected by the deformed planet varies with phase, thereby allowing the detection of deformation in high-precision light curves \citep{akin19, Hellard2019, Barros2022}. Since the projected area of the deformed planet varies throughout its orbit around the host star, it must also impact the observed orbital phase curve.

Despite many effects, including the stellar ellipsoidal variation, being accounted for when analyzing phase curve observations \citep[see e.g.,][]{Morris1985TheStars., Loeb2003PeriodicCompanions, Esteves2013OPTICALEXOPLANETS, Shporer2017-PCreview}, the planet shape is usually assumed to be spherical due to the expected low amplitude of tidal deformation \citep{Bourrier2020}. However, with the precision of present and upcoming instruments, the subtle effects of deformation may now be detectable in phase curves. Since the deformation of a planet depends on its interior structure, its detection will shed valuable insight into the composition of the planet \citep{Kellermann2018, Hellard2019, akin19}. Furthermore, at high precisions, disregarding the effect of planetary deformation may bias interpretations of phase curves of ultra-hot Jupiters (UHJs).

In this paper, we drop the assumption of sphericity but instead model the deformed planet shape as a triaxial ellipsoid to investigate the contribution of planet tidal deformation to phase curve observations of ultra-hot Jupiters.  In Section\,\ref{sect:pc_model}, we model the shape of a deformed planet and describe a numerical and analytical approach to account for tidal deformation in phase curves of UHJs. In Section\,\ref{sect:detectability}, we characterize the signature of tidal deformation across the full planetary orbit. We also compare the detectability of deformation in transit and phase curve observations for several space-based observatories. In Section\,\ref{sect:core_mass}, we illustrate how the measurement of tidal deformation can be used to constrain the core mass fraction of a planet, while Section\,\ref{sect:par_bias} investigates biases in parameters derived from phase curve analysis when sphericity is assumed for a deformed planet. We summarize our main conclusions in Section\,\ref{sect:conclusions}.

\section{Modeling tidal deformation}
\label{sect:pc_model}
\subsection{Deformation and Love number}
The deformation of a planet in response to stellar tidal forces depends on its interior structure and can be parameterized by the second-degree fluid Love number for radial displacement \hf{}. In hydrostatic equilibrium,  the giant planet deforms as a fluid body and we have that \hf{}\,=\,1 + $k_{2f}$, where $k_{2f}$ is the second-degree Love number for potential which measures the distribution of mass within the planet\footnote{Calculation of the different Love numbers can be found in \citet{sabadini} and \citet{Padovan2018}.}  \citep{love, Kellermann2018}. Since the Love number \hf depends on the distribution of mass within the planet, it measures the degree of central condensation of a body, thereby allowing us to constrain the interior structure \citep{kramm11,kramm12}. The value of \hf ranges from 0 to 2.5, although certain non-equilibrium processes in the interior and nonlinear perturbative effects can result in values outside this range \citep{Ragozzine2009, Wahl2021TidalJupiters}. A body with a fully homogeneous mass distribution (constant density) will have the maximum \hf value of 2.5 whereas a highly differentiated body with most of its mass condensed in a core will have a lower value (\hf{}$\simeq1$ for main sequence stars; \citealt{Ragozzine2009}). More massive objects tend to have smaller \hf values since they are more compressible and thus more centrally condensed \citep{Leconte2011}. Measurements of \hf can therefore be used to infer the presence and mass of a planetary core. For instance, Saturn's lower \hf of 1.39 \citep{LAINEY2017} is indicative of a higher core mass fraction than Jupiter with \hf{}=1.565 \citep{Durante2020}. Similarly, limits on the Love number of HAT-P-13\,b, obtained from its eccentricity due to its unique orbital configuration with a highly eccentric outer planet, allowed to place constraints on the maximum core mass of the planet \citep[][]{Batygin2009, kramm12,Buhler2016, hardy_2017}. However, these works derive different limits on the Love number of the planet, highlighting the difficulty in accurately measuring the Love number of exoplanets.

Therefore, measuring the shape of a tidally deformed giant planet from  its light curve allows estimating \hf which, in turn, provides direct constraints on the interior structure of the planet.

\subsection{Shape model}
\label{sect:shape_model}
The shape of a deformed planet is described here as a triaxial ellipsoid with semi-axes [$r_1,\,r_2,\,r_3$] where $r_1$ is the planet radius oriented along the star-planet (sub-stellar) axis, $r_2$ is along the orbital direction (dawn-dusk axis), and $r_3$ is along the polar axis. The volumetric radius of the ellipsoid is given by \Rv{}$~=~(r_1\,r_2\,r_3)^{1/3}$. According to \citet{correia14}, the axes of the ellipsoid are related by $r_2\,=\,\Rv(1-2q/3)$, $r_1$\,=\,$r_2(1+3q)$, and $r_3$\,=\,$r_2(1-q)$, where $q$ is an asymmetry parameter given by
\begin{equation}
    q  = \frac{\hf}{2Q_{\mathrm{M}}}\left(\frac{\Rv}{a}\right)^3\,
    \label{eqn:asym_par}
\end{equation}
and depends on the Love number \hf,  the planet-to-star mass ratio $Q_{\mathrm{M}}=M_p/M_{\star}$, and the orbital semi-major axis $a$. 

The non-spherical shape of the ellipsoidal planet implies that it projects a varying cross-sectional area as it rotates with orbital phase. The projected area as a function of orbital phase angle ($\phi$\,=\,2$\pi\,\times\,$phase) is given \citep[e.g. by][]{Leconte2011a} as 

\begin{equation}
    A(\phi) = \pi\sqrt{r_1^2r_2^2\cos^2i + r_3^2\sin^2i\,\left(r_1^2\sin^2{\phi} + r_2^2\cos^2{\phi}\right)}
\label{eqn:varying_area}
\end{equation}

\noindent where $i$ is the orbital inclination of the planet. The projected area of the ellipsoid is plotted in Fig.\,\ref{fig:phase_components}a for a planet with the size, mass, and \hf of Jupiter orbiting a Sun-like star at a distance of 3 stellar radii. The area is normalized to the projected area at mid-transit. We see that the ellipsoid projects its minimum area at mid-transit and mid-eclipse (phases 0 and 0.5) while the maximum is at quadrature where the projected area is 9\% larger than at mid-transit. As shown in previous studies  \cite[e.g.,][]{correia14, akin19, Hellard2019}, the varying projected shape and area of the ellipsoidal planet modifies the transit light curve. This allows the measurement of the planet's shape and its Love number but requires high-precision transit data. Indeed, \citet{Barros2022} were only able to detect the deformation of WASP-103\,b by combining transit observations from several space-based instruments for improved precision. They obtained a $3\sigma$ significant measurement of \hf. \citet{Hellard2020} obtained a less significant measurement for WASP-121\,b using only few transit observations from HST. Therefore, significantly detecting tidal deformation from transit observations remains quite challenging. We can further extend the previous works by probing for the varying area of the deformed planet in full-orbit phase curve observations, opening the potential for a more robust measurement of \hf that facilitates improved modeling of planetary interior structures.

\subsection{Phase curve model}

The phase curve model is a combination of planetary and stellar signals which we summarize in this section with a modification introduced to account for planet deformation. The phase curve model is composed of transit $F_{\mathrm{tra}}$, occultation $F_{\mathrm{occ}}$, planet atmospheric phase variation $F_p$, and stellar phase variation $F_{\star}$ signals and is given as a function of $\phi$ as:
\begin{equation}
F(\phi)  = F_{\mathrm{tra}}\times F_{\star}(\phi)~+~F_{\mathrm{occ}}\times F_p(\phi),\label{eqn:pc_flux}
\end{equation}

\noindent where the stellar phase variation is composed of the stellar ellipsoidal variation $F_{\mathrm{EV}}$ \citep{Morris1985TheStars.} and Doppler boosting $F_{\mathrm{DB}}$ \citep{Loeb2003PeriodicCompanions} signals as:  
\begin{equation}
    F_{\star}(\phi) = 1  + F_{\mathrm{EV}} + F_{\mathrm{DB}}.
    \label{eqn:st_flux}
 \end{equation}

\noindent The components of Eqs.\,\ref{eqn:pc_flux} and \ref{eqn:st_flux} are described in the following sections.

\subsubsection{Transit and Occultation}
\label{sect:tra_occ}
The planet's transit in front of the star and the occultation (secondary eclipse) behind the star can be modeled by several available tools. Here, we use the publicly available \texttt{ellc} transit tool \citep{maxted18} where our transit model $F_{\mathrm{tra}}$ adopts the power-2 limb darkening law \citep{Hestroffer1997} with $c_{\mathrm{LD}}$ and $\alpha_{\mathrm{LD}}$ as limb darkening coefficients (LDCs). The power-2 law has been shown to outperform other two-parameter laws in modeling intensity profiles generated by stellar atmosphere models \citep{morello, Claret_southworth2022_power2_ldcs}. The occultation model, $F_{\mathrm{occ}}$, is the same as the transit but without limb darkening and normalized to a depth of unity.

To model the transit signal of a deformed planet,  we follow the implementation of \citet{akin19} that incorporates the \citet{correia14} ellipsoidal shape model ($\S$\,\ref{sect:shape_model}) into \texttt{ellc}. Therefore, our generated transit light curve takes into account the varying shape and cross-sectional area of the ellipsoid during transit. In addition to the usual transit parameters, the ellipsoidal transit model takes  \hf and $Q_M$ as inputs while replacing the typical spherical planet radius $R_p$ by the volumetric radius \Rv (Eq.\,\ref{eqn:asym_par}). Spherical planet transit and occultation signals can also be generated by setting \hf{}$=0$, which makes \Rv{}\,=\,$R_p$.

\subsubsection{Ellipsoidal variation}
As previously mentioned, an ellipsoidal variation (EV) signal can be observed in phase curves of UHJs as a result of stellar deformation by a massive planet \citep{Morris1985TheStars.,mislis2012}. Detailed expressions for the flux variations due to EV can be found in \citet{Esteves2013OPTICALEXOPLANETS} and \citet{csizmadia_wavelets2023}. However, the EV signal can be approximated by a cosine function at the first harmonic of the orbital period \citep{Shporer2017-PCreview}, so it has two peaks (at phases 0.25 and 0.75) during an orbit. It is given as
\begin{equation}
    F_{\mathrm{EV}} = A_{\mathrm{EV}}(1 - \cos{2\phi}),
\end{equation}
where the semi-amplitude $A_{\mathrm{EV}}$ depends on the planet-to-star mass ratio $Q_{\mathrm{M}}$, the scaled semi-major axis $a/R_{\ast}$, and the inclination ${i}$ as 
\begin{equation}
    A_{\mathrm{EV}} = \alpha_{\mathrm{EV}}\,Q_{\mathrm{M}}\left(\frac{R_{\ast}}{a}\right)^3\sin^2i
    \label{eqn:EV}
\end{equation}
where $\alpha_{\mathrm{EV}}$ is a factor that depends on the linear limb-darkening coefficient and gravity-darkening coefficient.

\begin{figure}
    \centering
    \includegraphics[width=\linewidth]{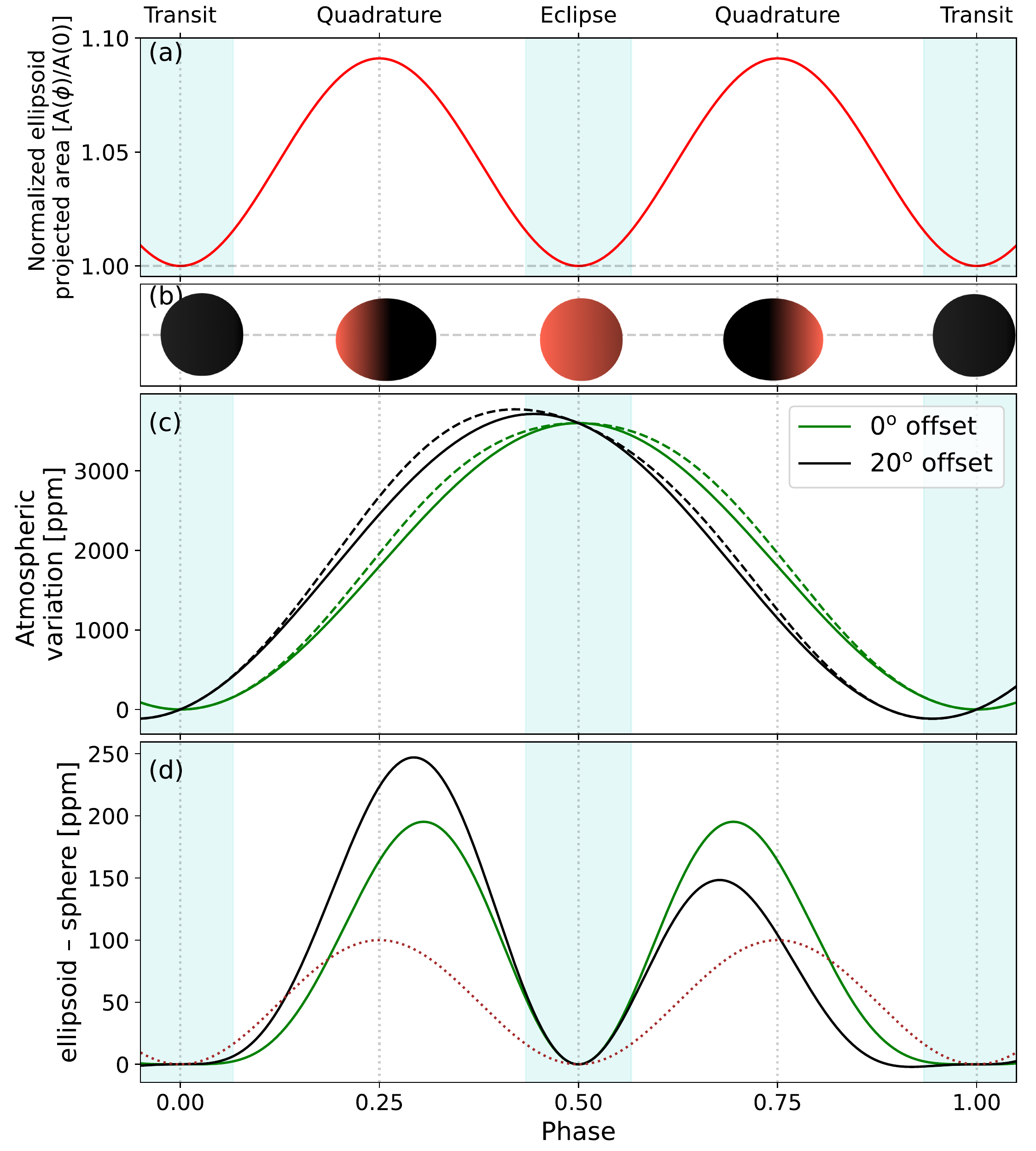}
    \caption{Deformed planet phase curve components. \textbf{(a)}~Phase-dependent projected area of an ellipsoid normalized to the projected area at phase zero (mid-transit). \textbf{(b)}~Schematic of the projected ellipsoidal planet shape at different phases and the illuminated regions. \textbf{(c)}~Atmospheric phase variation of a spherical (solid) and ellipsoidal (dashed) planet calculated analytically using Eq.\,\ref{eqn:def_Fp}. The case with (green) and without (black) phase offset is shown. \textbf{(d)}~Difference between the ellipsoidal and spherical planet atmospheric phase variations for the cases with and without phase offset. This represents the contribution of deformation to the planetary signal. For comparison, the stellar ellipsoidal variation $F_{\mathrm{EV}}$ with a peak-to-peak amplitude of 100\,ppm is also shown as the brown dotted curve. The cyan-shaded regions in all panels represent the in-transit and in-eclipse phases.}
    \label{fig:phase_components}
\end{figure}

\subsubsection{Doppler beaming}
This is an orbital photometric modulation caused by the motion of the star around the system's center of mass. The motion results in a Doppler shift of stellar light leading to a variation in the amount of photons observed in a particular passband \citep{Loeb2003PeriodicCompanions, Shporer2017-PCreview}. The signal is described by a sine curve at the orbital period so it has one peak (at phase 0.25) during an orbit. It is given by
\begin{equation}
    F_{\mathrm{DB}} = A_{\mathrm{DB}} \sin{(\phi)}
\end{equation}
where the semi-amplitude $A_{\mathrm{DB}}$ is given by 
\begin{equation}
    A_{\mathrm{DB}} = \alpha_{\mathrm{DB}}4\frac{K}{c}.
    \label{eqn:DB}
\end{equation}
Here $K$ is the RV semi-amplitude, $c$ is the speed of light, and $\alpha_{\mathrm{DB}}$ is a coefficient of order unity that depends on the target's spectrum in the observed passband. For most planets with sub\,–\,km$\,s^{-1}$ RV amplitudes, $A_{\mathrm{DB}}$ is very small ($<$5\,ppm) and so Doppler beaming has little effect on the phase curve signal.

\subsubsection{Planetary atmospheric phase variation}
As a planet orbits its star, the planetary flux emanating from different fractions of the projected hemisphere contributes to the total observed flux from the system, resulting in a variation of the observed flux with phase. The planetary flux is composed of reflected light and thermal emission from the atmosphere but these components are difficult to disentangle in observations as they produce variations that are similar in shape. However, since planets susceptible to tidal deformation are strongly irradiated, most of the observed flux is expected to be from thermal emission rather than reflected light. Here, we calculate the total planetary phase variation using numerical and analytical methods in order to investigate the contribution of planetary tidal deformation.

\subsubsection*{Numerical calculation}

\begin{figure*}[ht]
    \centering
    \includegraphics[width=\linewidth]{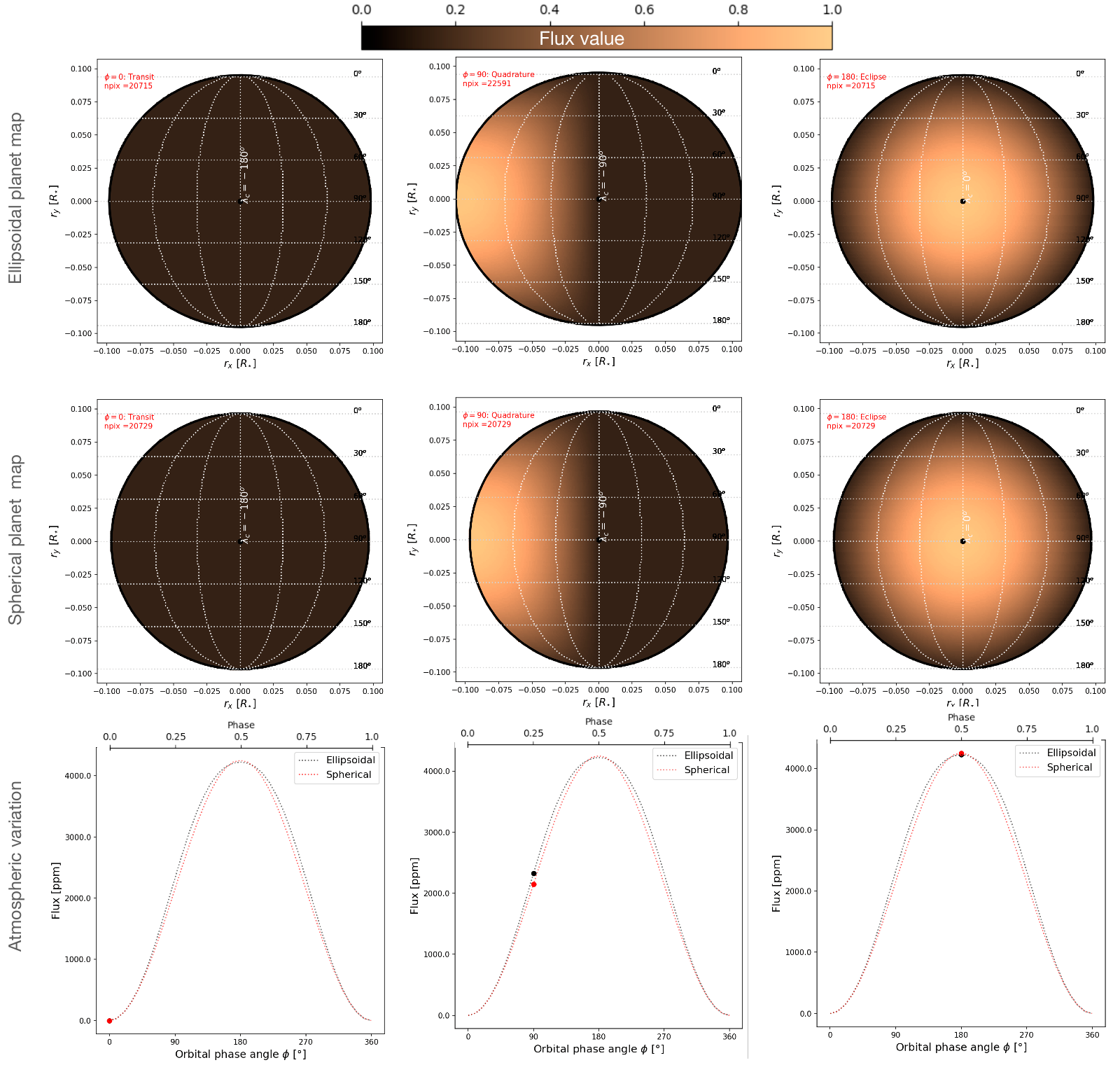}
    \caption{Schematic of numerical calculation of planetary phase variation for an ellipsoidal and spherical planet. \textbf{Top:} Planet flux map for an ellipsoidal planet at transit (left), quadrature (middle), and eclipse (right) phases. Note the changing projected area of the planet with the number of pixels at each phase given as $npix$.  \textbf{Middle:} Same as top panels but for a spherical planet. \textbf{Bottom:} Planet flux variation obtained from integrating the flux map at the different phases.  The black and red circles indicate the integrated flux value at each illustrated phase for the ellipsoidal and spherical planets, respectively.}
    \label{fig:num_pc}
\end{figure*}

To numerically simulate the planetary phase variation signal, we adopt a pixelation approach where the planet (ellipsoidal or spherical) is projected on a 2-dimensional Cartesian grid ($XY$). The projected shape of the ellipsoidal planet is an ellipse with semi-axes, $r_x$ and $r_y$ on the grid, that varies with orbital phase and can be derived from Eq.\,\ref{eqn:varying_area}. On the other hand, the projected shape of the spherical planet is a circle with radius, $r_x$=$r_y$, that remains constant with orbital phase.

At each phase, a flux map is assigned to the projected planet. Different flux maps can be adopted \citep[see e.g.,][and references therein]{Louden2018SPIDERMAN:Eclipses}. Here, we define a flux map given as a function of co-latitude ($\theta$) and longitude ($\lambda$) of the visible hemisphere as:
\begin{equation}
    \mathcal{F}(\theta,\lambda)=A_0 + A_1 \sin \theta \cos{(\lambda + \delta)},
\end{equation}

\noindent which simulates the simple sinusoidal phase function typically used to model the total brightness variation of planets \citep[e.g.,][]{cowan2012-wasp-12,kreid18, shporer-2019-WASP18, bell-2019-WASP-12,wong2021_tess2ndyearPCs,Zieba2022K2Atmosphere}. Here, $\theta$ varies from $0-180\degree$ between the poles, while $\lambda$ varies from $\lambda_c\,-\,90\degree$ to $\lambda_c\,+\,90\degree$ with $\lambda_c$ denoting the central longitude of the planet at a particular orbital phase (see Fig.\,\ref{fig:num_pc}). The longitudes are defined such that the substellar point is on $\lambda=0\degree$ and the antistellar point is on $\lambda=180\degree$. The hotspot offset is denoted by $\delta$. The coefficients $A_0$ and $A_1$ define the flux value of pixels at the dawn/dusk limbs of the planet, and the flux difference between these limbs and the center at eclipse, respectively. For illustration, we set $A_0=0$ to simulate a total nightside flux of zero, while we set $A_1=1$ so as to have a flux value of 1 at the substellar point. The planet's phase variation $F_p(\phi)$ can then be obtained by integrating the flux map across the projected surface at each phase.

Assuming a planet with the size, mass, and \hf of Jupiter with $a/R_{\star}\,=\,3$, Fig.\,\ref{fig:num_pc} shows the flux map for the ellipsoidal and spherical planet cases at transit, quadrature, and eclipse phases. Notice the larger projected area of the ellipsoidal planet at quadrature where there are more "brightly emitting" pixels compared to the spherical planet. The generated phase variation signals are also shown, where we see that the varying projected area of the ellipsoidal planet leads to additional flux between transit and eclipse since the emitting surface is larger at these phases. The maximum difference between the ellipsoidal and spherical planet occurs slightly after quadrature where the observer sees the largest illuminated/emitting surface of the ellipsoidal planet. Meanwhile, by construction, the spherical planet signal is equivalent to a cosine function.

Numerically computing the planetary contribution to the phase curve can be computationally intensive, especially when fitting to observations. Therefore, we consider an analytical solution.

\subsubsection*{Analytical calculation}

Following previous phase curve studies \citep[e.g.,][]{cowan2012-wasp-12,shporer-2019-WASP18, wong2021_tess2ndyearPCs}, we analytically model the total atmospheric phase variation with a cosine function between the minimum flux, $F_{\mathrm{min}}$, and maximum flux $F_{\mathrm{max}}$, as
\begin{equation}
    F_{p}(\phi)  =  \left(F_{\mathrm{max}} - F_{\mathrm{min}}\right) \, \frac{1 - \cos{(\phi + \delta)}}{2} + F_{\mathrm{min}}.
    \label{eqn:planet_flux}
\end{equation}
\noindent The dayside flux $F_d$ (i.e., occultation depth) and nightside flux $F_n$ of the planet are derived as the value of $F_p(\phi)$ at $\phi=\pi$ and $\phi=0$ respectively, while the semi-amplitude of the atmospheric variation $A_{\mathrm{atm}}$ is $(F_{\mathrm{max}} -  F_{\mathrm{min}})/2$.

As seen in the numerical simulation, the main effect introduced by tidal deformation in phase curves is geometric, owing to the changing projected area of the ellipsoid. Therefore, we account for tidal deformation by multiplying the planet's phase variation (Eq.\,\ref{eqn:planet_flux}) by the normalized phase-dependent projected area of the ellipsoid\footnote{Note that this is done only for the out-of-transit/occultation phases since the projected shape and area of the deformed planet is already accounted for in the transit/occultation models (\S\,\ref{sect:tra_occ}) } (Eq.\,\ref{eqn:varying_area}): 
\begin{equation}
    F_p(\phi)^{\mathrm{def}}= F_p(\phi)\,\frac{A(\phi)}{A(0)}.
    \label{eqn:def_Fp}
\end{equation}

\noindent We note that other phase variation models (e.g. Kinematic temperature model of \citealt{Zhang2017EffectsExoplanets}, or other analytical functions as given in \citealt{csizmadia_wavelets2023}) can be adopted for $F_p(\phi)$ and the same operation can be performed to incorporate tidal deformation. The analytical model produces an equivalent planetary phase variation signal to that obtained numerically (see Fig.\,\ref{fig:comp_num_analy}), thereby validating the simplified  model which we use hereafter.

Figure\,\ref{fig:phase_components}c shows the atmospheric phase variations with zero phase offset for a spherical planet (green solid curve) and an ellipsoidal planet (green dashed curve) both with $F_d$\,=\,3600\,ppm\footnote{ similar to \spitzer 3.6\,$\mu$m and 4.5$\mu$m occultation depth measurements for a planet with a dayside temperature of $\sim$2700\,K orbiting a Sun-like star.}. The phase variation signals with eastward hotspot offset of 20$\degree$ (black curves) are also shown. Fig.\,\ref{fig:phase_components}d shows the difference between the phase variations of the ellipsoidal and spherical planets  (i.e. $F_p(\phi)^{\mathrm{def}} - F_p(\phi)$), highlighting the contribution of tidal deformation. In the case without hotspot offset (green curve), the contribution is symmetric about the mid-eclipse phase, peaking between quadrature and eclipse on both sides with an amplitude of $\sim$200\,ppm. In the case of a hotspot offset (black curve), the deformation contribution is asymmetric with some of its power shifted to the peak before eclipse giving a higher amplitude of $\sim$250\,ppm before eclipse and 150\,ppm after eclipse. The plot also shows the stellar ellipsoidal variation signal $F_{\mathrm{EV}}$ with an arbitrarily chosen peak-to-peak amplitude of 100\,ppm. The EV signal peaks at quadrature (i.e. spaced by 0.5 in phase) making it distinguishable from the planetary deformation signal which has peaks at different phases with shorter spacing between them.

\begin{figure}
    \centering
    \includegraphics[width=1\linewidth]{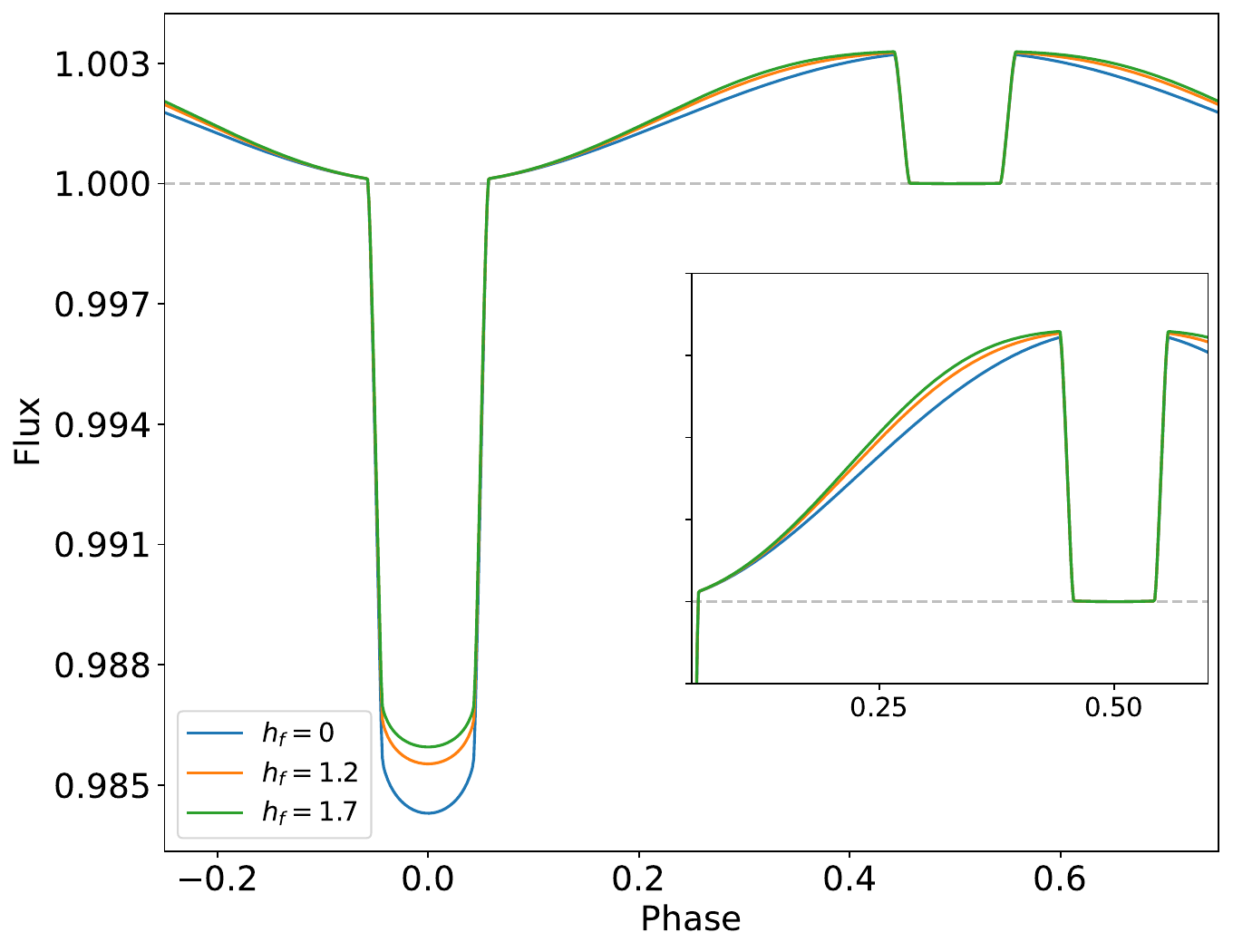}
    \caption{Ellipsoidal planet phase curves with different values of \hf, where \hf{}\,=\,0 is equivalent to a spherical planet. The inset zooms on the phase curve region after transit.}
    \label{fig:pc_diff_hf}
\end{figure}

\medskip
Our total phase curve implementation accounts for deformation by replacing $F_{\mathrm{tra}}$, $F_{\mathrm{occ}}$, and $F_p(\phi)$ in Eq.\,\ref{eqn:pc_flux} by their ellipsoidal planet equivalents. Fig.\,\ref{fig:pc_diff_hf} shows the ellipsoidal planet phase curve for different values of \hf with \hf=0 corresponding to a spherical planet. We see that the transit depth varies for different values of \hf since larger \hf implies more deformation which causes the ellipsoidal planet to project a smaller cross-section of its shape during transit. However, larger values of \hf increase the projected area of the ellipsoid towards quadrature, thereby causing an increase in the flux received from the planet between transit and occultation.

\section{Signature and detectability of tidal deformation}
\label{sect:detectability}

Tidal deformation is only significant for UHJs orbiting close to the Roche limit of their stars which limits the detectability to only a handful of confirmed planets. Previous studies \citep[e.g.,][]{correia14,akin19,Hellard2019,Berardo-Dewitt2022} have identified WASP-12\,b, WASP-103\,b, and WASP-121\,b as some of the most promising targets to measure deformation. For the analyses presented below, we adopt the parameters of WASP-12\,b as representative of such UHJs.

\subsection{The signature of deformation}
The detectability of deformation requires that the observed light curve of a deformed planet is distinct such that it cannot be accurately reproduced by a spherical planet model. From visual inspection of Fig.\,\ref{fig:pc_diff_hf}, one can imagine that most of the differences between the spherical and deformed planet phase curves can perhaps be explained by a modification of some of the spherical planet model parameters (e.g., radius, hotspot offset and others). However, in actual observations, the value of these parameters would be initially unknown but determined by performing a fit to the data. Therefore, the detectable signature of deformation is the residuals obtained from fitting a deformed planet observation (transit or phase curve) with a spherical planet model. The amplitude and timescale of the residual anomalies indicate the prominence of deformation in the data and hence its detectability.

\begin{table}[t!]
\caption{Parameters adopted in our optical and NIR  phase curve simulations of WASP-12\,b.}
\label{tab:sim_wasp12}
\resizebox{1.03\linewidth}{!}{%

\begin{tabular}{llll}
\hline\hline
Parameter                                              & Description                 & \begin{tabular}[c]{@{}l@{}}Optical/\\ TESS$^\dagger$\end{tabular}                                                    & \begin{tabular}[c]{@{}l@{}}NIR/\\ PRISM$^\ddagger$\end{tabular}                                                   \\ \hline
$R_p\,[R_{\ast}]$                                      & Scaled spherical radius & 0.1160                                                  & 0.1168                                                  \\

$\Rv\,[R_{\ast}]$                                      & Scaled volumetric radius$^\lozenge$ & 0.1222                                                  & 0.1232                                                  \\
$a/R_{\ast}$                                           & Scaled semi-major axis      & 3.061                                                   & 3.061                                                   \\
$b$                                                    & Impact parameter            & 0.344                                                   & 0.344                                                   \\[0.2em]
\begin{tabular}[c]{@{}l@{}}$c_{\mathrm{LD}}$\\ $\alpha_{\mathrm{LD}}$\end{tabular} & Power-2 LDCs$^\S$                & \begin{tabular}[c]{@{}l@{}}0.6371\\ 0.6175\end{tabular} & \begin{tabular}[c]{@{}l@{}}0.2862\\ 0.4183\end{tabular} \\[0.8em]
$F_d/F_{\ast}$\,{[}ppm{]}                              & Dayside flux               & 466                                                     & 3306                                                    \\
$F_n/F_{\ast}$\,{[}ppm{]}                              & Nightside flux              & 0                                                       & 500                                                     \\
$\delta\,[\degree]$                                      & Phase offset                & 6.2                                                     & 20                                                      \\
$A_{\mathrm{EV}}$\,{[}ppm{]}                        & EV semi-amplitude           & 65                                                      & 60                                                      \\
$A_{\mathrm{DB}}$\,{[}ppm{]}                         & DB semi-amplitude           & 2.34                                                    & 1.18                                                    \\ \hline
\hf                                                  & Love number$^\lozenge$               & \multicolumn{2}{c}{1.565}                                                                                         \\
$Q_{\mathrm{M}}$                                       & Planet-to-star mass ratio$^\lozenge$ & \multicolumn{2}{c}{0.00098}   \\ \hline
\end{tabular}%
}
\\
\textbf{Notes:} $^\dagger$\,The \tess parameter values are the derived values from  the analysis of 4 sectors of WASP-12 \tess observations \citet{Wong2022-WASP-12}. $^\S$\,These values were obtained using \texttt{LDTk} \citep{Parviainen2018} in the different passbands.
$^\ddagger$\,Values for the PRISM parameters $R_p$ and $F_d/F_{\ast}$ are obtained by integrating the modeled transmission \citep{Stevenson2014TRANSMISSIONm} and emission spectra \citep{Hooton2019StormsWASP-12b} across the PRISM wavelength range, $A_{\mathrm{EV}}$ and $A_{\mathrm{DB}}$ are estimated using Eqs.\,\ref{eqn:EV} and \ref{eqn:DB} respectively, while $\delta$ is the derived value from the 2013 \spitzer 4.5$\mu$m phase curve of WASP-12\.b \citep{bell-2019-WASP-12}.
$^\lozenge$\,These parameters are specific to the ellipsoidal planet model where we assume the same \hf as Jupiter, \Rv is the volumetric radius that produces the same transit depth as the spherical planet of radius $R_p$, and $Q_M$ is taken from \cite{Collins2017TRANSITPLANETS}.
\end{table}

To estimate and compare the signature of deformation in transit and phase curve observations, we assume broadband observations of WASP-12\,b in the optical (e.g. with \tess, \cheops, or \plato) and band-integrated observations in the near-infrared (NIR; e.g., with \jwst or \hst). We compared the deformation signatures by simulating ideal transit and phase curve observations (no uncertainties) of a deformed planet and performing a least-squares fit\footnote{with  \href{https://lmfit.github.io/lmfit-py/}{\texttt{LMFIT}} \citep{Newville2020Lmfit/lmfit-py1.0.1}} using a spherical planet model. We selected \tess for the representative simulation of the optical observations based on the available measurements and \jwst/\nirspec in PRISM mode for the NIR. Although we used \tess and \nirspec-PRISM as examples of observing instruments, the approach is applicable to other instruments observing in the optical and NIR. The adopted parameters of the simulation are given in Table\,\ref{tab:sim_wasp12} and they are all allowed to vary freely except for the LDCs and $A_{\mathrm{DB}}$ for which we used Gaussian priors. In the transit case, the out-of-transit baseline is kept short to minimize the effect of orbital phase variation on the fit and we put strong priors on the parameters ($F_d$, $F_n$, $\delta$, $A_{\mathrm{EV}}$) since they cannot be accurately estimated from the transit fit. For real observations, the phase variation in transit-only observations can be accounted for either by applying priors on the aforementioned parameters from previous phase curve analyses/observations of the target or by detrending using a time-dependent function.

\begin{figure}
    \centering
    \includegraphics[width=\linewidth]{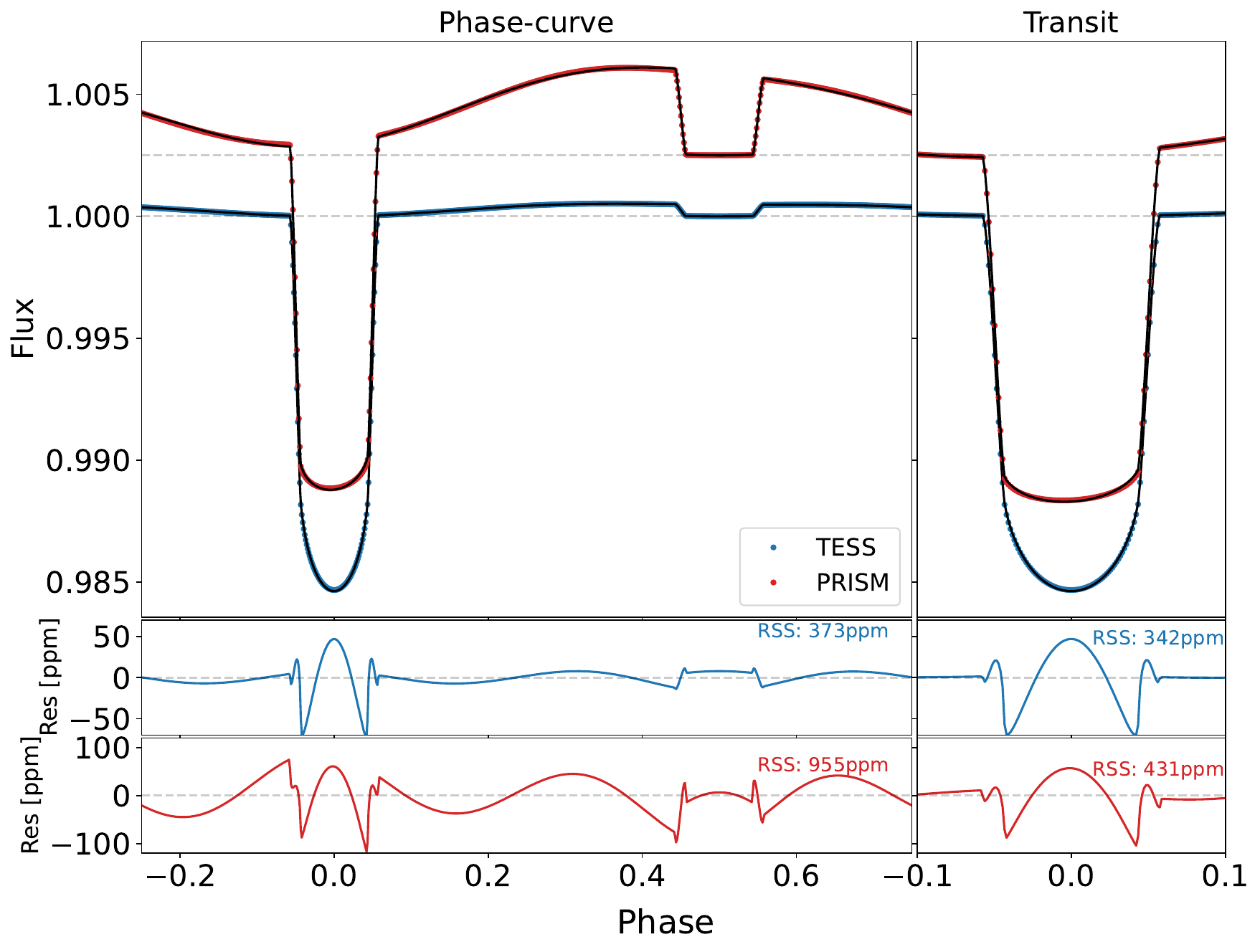}
    \caption{Fitting the phase curve and transit of a deformed planet with a spherical planet model. \textbf{Left:} Simulated deformed planet phase curve of WASP-12\,b in the \tess (blue) and \nirspec–PRISM (red) bands. The spherical planet model fit is overplotted for each light curve. The bottom panel shows the residuals of the fit for each light curve. \textbf{Right:} Same as the left panel but for observation of the transit only. The residuals are shown in the bottom for each light curve with the root-sum-of-squares (RSS) quoted.}
    \label{fig:def_signature}
\end{figure}

The results of the fits are shown in Fig.\,\ref{fig:def_signature}. The residuals of the transit fits display anomalies that represent the signature of deformation in transit light curves as described in \citet{correia14} and \citet{akin19} – transit ingress and egress anomalies due to the longer transit of the ellipsoidal planet, and in-transit anomaly due to varying projected shape and eclipsed area by the rotating ellipsoidal planet. The residuals of the phase curve fits show an additional anomaly across the orbit also due to the varying projected area of the ellipsoidal planet, which contributes additional flux at different phases that cannot be easily accounted for by the spherical planet model. There are also ingress and egress anomalies in the occultation due to the longer occultation duration of the ellipsoidal planet.

\begin{figure*}[th!]
    \centering
    \includegraphics[width=0.495\textwidth]{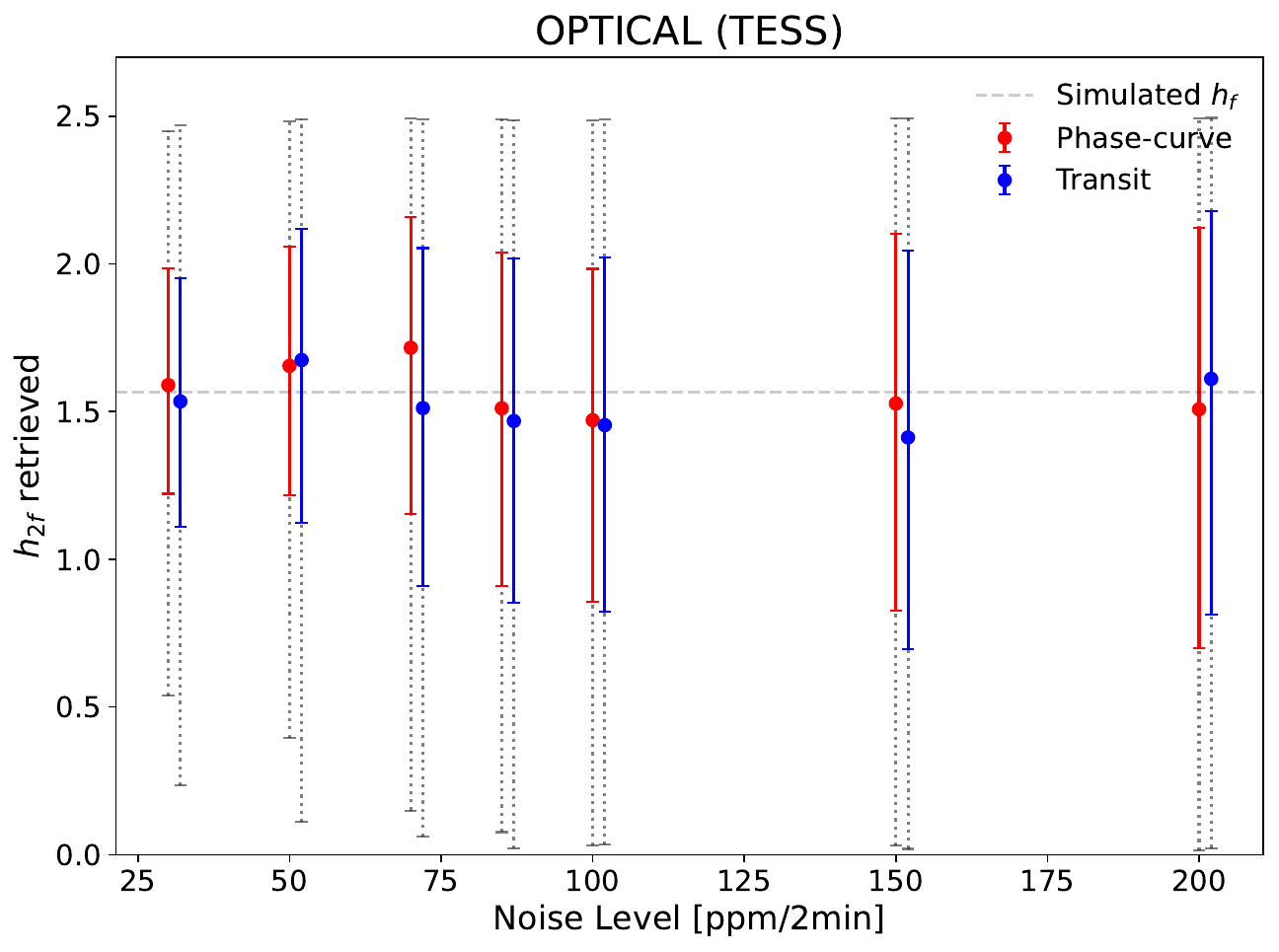}
    \includegraphics[width=0.495\textwidth]{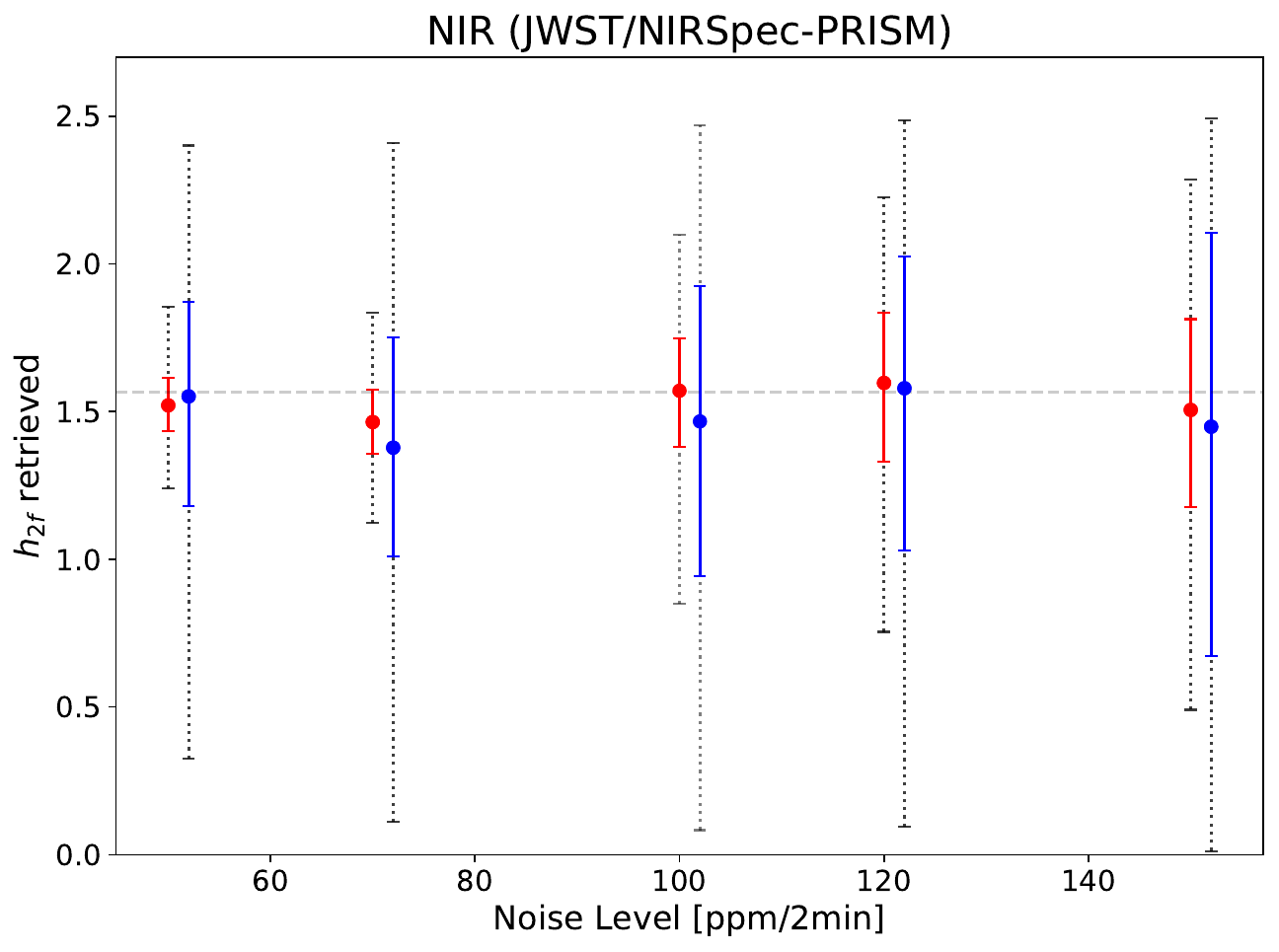}
    \caption{Detectability of deformation in the optical (\tess; left) and NIR (\jwst/\nirspec PRISM; right). The horizontal dashed line indicates the simulated value of \hf=1.565. The red points show the retrieved value of \hf from phase curves and the 1$\sigma$ uncertainty while the blue points show the retrieved \hf from transits. The gray dotted errorbars show the 3$\sigma$ uncertainty of the measurements. A similar plot for \hst/\wfc is shown in Fig.\,\ref{fig:pc_fit_hst}. }
    \label{fig:PCfit}
\end{figure*}

To quantify the deviation of the spherical model fits from the deformed planet simulated observations, we computed the root-sum-of-squares (\textit{RSS}) of the residuals. We find a greater deviation in the PRISM transit residuals than for \tess indicating a larger deformation signal in NIR than optical due to the reduced effect of limb darkening in the NIR that amplifies the ingress and egress anomalies. The phase curve residuals also exhibit larger deviations than the corresponding transit residuals ($>$2$\times$ for PRISM) due to the wider phase range of the phase curve deformation signature. This implies that the effect of deformation is more prominent in phase curves than in transit-only observations. Comparing the phase curve deformation signature between the instruments, we see that the out-of-transit amplitude is just as high as the in-transit amplitude for PRISM, whereas it is much smaller for \tess. Indeed, the \textit{RSS} of the TESS phase curve signature does not provide much improvement compared to the \textit{RSS} of the transit due to the small contribution from the out-of-transit signature. 

In summary, phase curves provide a longer phase range than transits which allows to better probe the shape of the planet as it rotates with phase. The longer phase range can also be more easily sampled to attain better precision that facilitates the detection of the anomaly.  We quantify the detectability of deformation from transits and phase curves in the next section.

\subsection{Detectability of tidal deformation in transit and phase curve observations}
\label{sect:detectability_comp}

Detecting tidal deformation implies measuring a statistically significant value of \hf from fitting transit and/or phase curve observations with the ellipsoidal planet model described in Section\,\ref{sect:pc_model}. We investigate the detectability of tidal deformation at optical and NIR passbands by performing injection-retrieval simulations. We simulate the transit and phase curve of deformed WASP-12\,b at a cadence of 2 minutes using the parameters listed in Table\,\ref{tab:sim_wasp12}.  For ease of fitting, we do not include any phase variation in the simulated transit signals, assuming that this will be accounted for as previously mentioned. 

In order to investigate the photometric precision required to detect deformation and assess the detectability by different instruments, we added random Gaussian noise of different levels to the simulated transit and phase curve which we then fit with ellipsoidal and spherical planet models. The \texttt{dynesty} python package \citep{speagle} was used to sample the parameter space and derive parameter posteriors. All parameters concerning each model were allowed to vary freely except for the power-2 LDCs for which we used Gaussian priors with arbitrary uncertainties of 0.01, $Q_M$ whose prior was based on derived values in \cite{Collins2017TRANSITPLANETS}, and $A_{\mathrm{DB}}$ whose prior was derived from Eq.\,\ref{eqn:DB}.

For each noise level, we generated 20 simulated observations by adding random realizations of the noise. We then performed the model fit to each observation. The posterior distributions from the fits with the different noise realizations are then merged. This method marginalizes over several possible noise realizations and thus mitigates the impact of obtaining biased results based on a single randomly generated favorable/unfavorable noise realization.

The detectability of tidal deformation in the optical and NIR is summarized in Fig.\,\ref{fig:PCfit}. It shows the retrieved value of \hf from the fits to the simulated optical and NIR observations at different noise levels. For the optical simulations, the transit fits allow recovering \hf with 3$\sigma$ significance for noise levels up to 70\,ppm/2min, while the phase curve fits extends this to a higher noise level of 85\,ppm/2min. At noise levels $\leq$50\,ppm/2min, the phase curve fits provide up to 15\% improvement in the precision of retrieved \hf compared to the transit fits. However, the improvement decreases below 5\% at higher noise levels since the relatively small amplitude of the deformation signature outside of transit phases becomes indiscernible from the noise. For the NIR simulations, \hf is recovered at 3$\sigma$ in the transit fits for noise levels up to 120\,ppm/2min. Across all added noise levels, the phase curve fits provide a significant improvement (76\% at 50\,ppm and $>$30\% at 150\,ppm) in the precision of retrieved \hf compared to the transit fits due to the long phase range and large amplitude deformation signature outside transit. 

Our injection and retrieval simulations thus imply that detecting tidal deformation is more favorable at NIR passbands compared to the optical since we are able to measure \hf significantly at higher noise levels in the NIR, particularly for phase curve observations. 

As phase curves require more observing time than transits, we compare \hf retrieval from phase curves to the same observing time in transit observations. This allows to determine the most effective observing strategy for the precise measurement of \hf. For the same observing time, we find that phase curves provide improved \hf precision over the combined transit by up to 10\% in the optical and up to 50\% for \jwst.

 \subsubsection{Impact of limb-darkening on detectability}
The effect of limb darkening is most significant at ingress and egress, where the signature of tidal deformation also manifests strongly. Limb darkening is therefore capable of compensating for part of the deformation-induced signal, thus reducing detectability, particularly for transit observations \citep{Hellard2019,akin19}. Therefore, detecting deformation requires proper treatment of limb darkening. 

We find in our injection–retrieval simulations that increasing the width of our limb darkening priors results in reduced detection significance. For instance, when the widths of the LDC priors are slightly increased from 0.01 to 0.015 in the optical simulations, the noise level required for a 3$\sigma$ detection reduces from 70 to 50\,ppm/2min in the transit fits and from 85 to 55\,ppm/2min in the phase curve fits. The reduction in both cases is because most of the detection power in the optical comes from the transit phases where limb darkening acts. However, the increased LDC prior width in the NIR causes only a slight reduction in noise level from 120 to 110\,ppm/2min for a 3$\sigma$ transit detection due to the weaker limb darkening effect in the infrared. The NIR phase curve fits remain largely unchanged, even with further increase in the LDC prior width, since the deformation signature spans a wide phase range with a large out-of-transit amplitude that allows a precise constraint of the deformation inspite of limb darkening.

\subsubsection{Disentangling deformation from atmospheric phenomena}
It is possible for other atmospheric effects to confound the deformation signal or make its detection challenging. For example, highly irradiated planets may exhibit strong day-to-night temperature gradients that can result in different chemical compositions and scale heights between the two hemispheres \citep[see e.g.,][]{Pluriel2020StrongHeterogeneities,Helling2020MineralWASP-43b}. This leads to limb asymmetries \citep{Espinoza2021ConstrainingSpectroscopy} that can affect the detectability of deformation in transit. 

To investigate whether the effects of day-to-night inhomogeneities can mimic tidal deformation, we simulated the transit of a planet with limb asymmetries using \texttt{catwoman} \citep{Jones2020Catwoman:Curves} and fit it with a deformed planet model. We find that tidal deformation is unable to account for limb asymmetries since tidal deformation does not cause asymmetric transits. Conversely, we also found that \texttt{catwoman} is unable to accurately fit a deformed planet transit indicating that the two effects are distinguishable.

Furthermore, the effect of day-night side inhomogeneities is chromatic and is dependent on the composition of the day and nightside. In contrast, the Love number is an intrinsic property of the planet that remains constant across wavelengths. Therefore, we can differentiate between them by ensuring that consistent \hf is derived across different spectral bins or passbands. Therefore, spectroscopic observations with \hst and \jwst will be very useful in this regard.

\begin{figure*}[ht]
    \centering
    \includegraphics[width=0.8\linewidth]{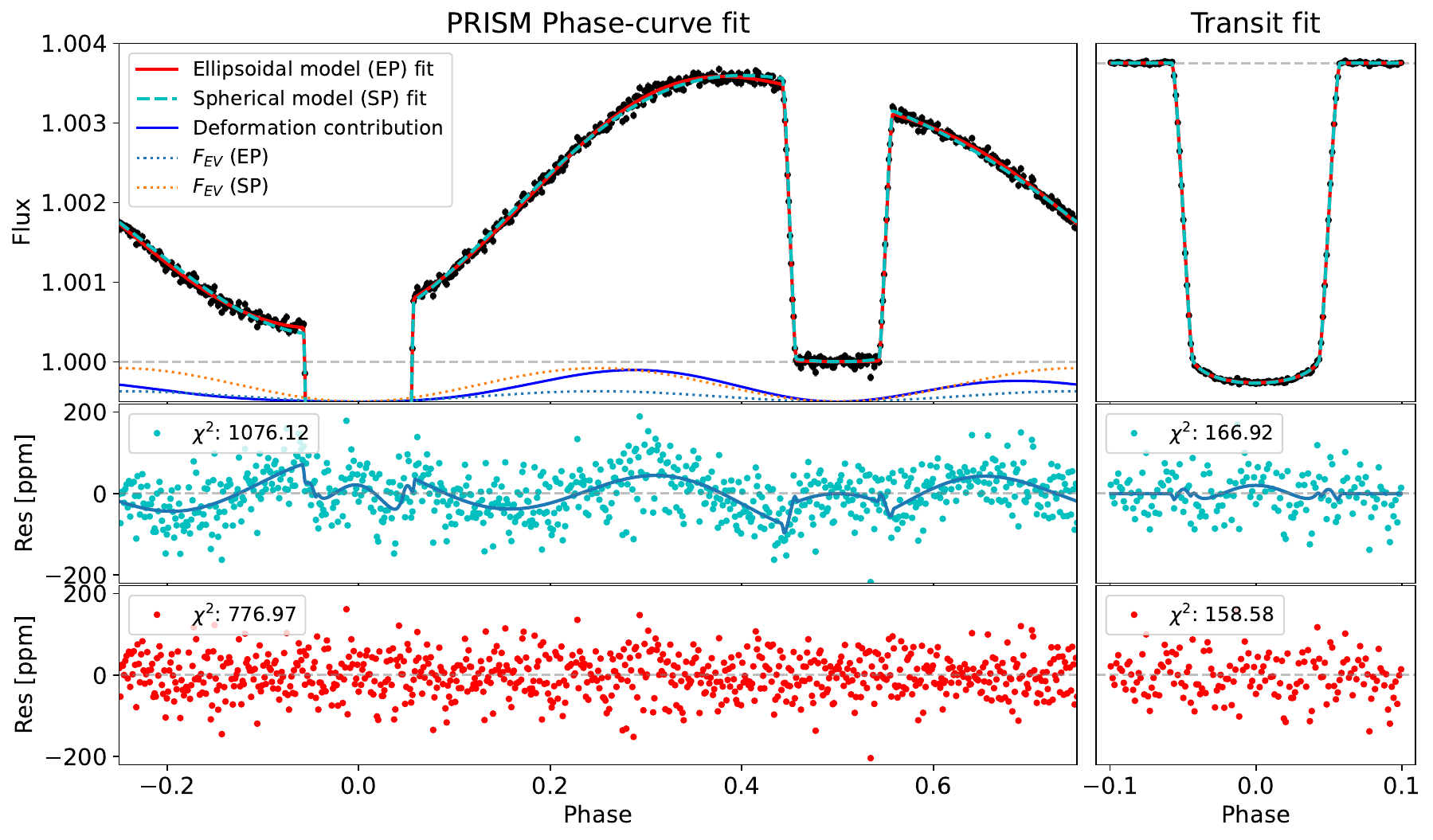}
    \caption{Spherical planet (SP) and ellipsoidal planet (EP) model fits to the simulated deformed planet signals with a PRISM noise level of 50\,ppm/2min. The \textit{top} panel shows the simulated phase curve (left) and transit (right) observations with the fitted models overplotted. The phase curve fit also shows the contribution of deformation to the light curve (blue) and the $F_{\mathrm{EV}}$ component of the EP and SP model fits. The \textit{middle} and \textit{bottom} panels show the residuals from fitting a spherical planet model (cyan) and an ellipsoidal planet model to the simulations.}
    \label{fig:jwst_pcfit_50ppm}
\end{figure*}

\subsubsection{Detectability of deformation by different instruments}
\label{sect:compare_dect}

 Figure\,\ref{fig:PCfit} can be used to estimate the detectability of deformation by different instruments observing phase curves or transits of close-in giant planets in the optical and NIR. Here, we estimate the detectability for observations with \jwst, \plato, \tess, \cheops, and \hst. 
 
\medskip
\noindent\textbullet \, \textbf{\jwst}: Using the \jwst Exposure Time Calculator\footnote{\href{https://jwst.etc.stsci.edu/}{https://jwst.etc.stsci.edu/}}, we estimated a precision of 40\,–\,50\,ppm/2min for observations of WASP-12\,b with different \jwst instruments, for example \nirspec-PRISM. At the similar noise level of 50\,ppm/2min, our NIR simulations recovered a highly significant detection of \hf{}$=1.52^{+0.094}_{-0.087}$ (17$\sigma$) from the phase curve fit and \hf{}$=1.55^{+0.32}_{-0.37}$ (4.2$\sigma$) from the transit fit. Fig.\,\ref{fig:jwst_pcfit_50ppm} shows the ellipsoidal and spherical planet model fits to the simulated PRISM observations with one realization of the 50\,ppm noise. The residual of the spherical planet fit shows a clear modulation induced by tidal deformation that cannot be accounted for by the spherical planet. Fig.\,\ref{fig:jwst_pcfit_50ppm} further shows the contribution of planet deformation (cf. Fig.\,\ref{fig:phase_components}d) to the total observed phase curve in comparison to the best-fit stellar ellipsoidal variation signals $F_{\mathrm{EV}}$. We observe that the amplitude of $F_{\mathrm{EV}}$ in the spherical planet model fit increases (from the simulated value of 60\,ppm to 210\,ppm) in an effort to account for the deformation signal of the planet. However, it is unable to do so since the deformation contribution peaks at different phases with a shorter inter-peak interval compared to $F_{\mathrm{EV}}$. This is still the case when a tidal phase lag parameter is added to $F_{\mathrm{EV}}$. This implies that the assumption of sphericity for a deformed planet can lead to biases in the estimation of other system parameters. In Section\,\ref{sect:par_bias}, we further investigate the bias on different system parameters when assuming sphericity for a deformed planet.

\medskip
\noindent\textbullet \, \textbf{\plato}: For stars brighter than V$\simeq$11, \plato will be capable of achieving a photometric precision of 34\,ppm/hr \citep[][]{rauer,Rauer2018}  equivalent to 186ppm/2min. The observation strategy of \plato consists of a long-pointing phase where two different fields are observed for 2 – 3 yrs each, and a short observation (step and stare) phase where each stare would last 2 – 3 months \citep{Rauer2018}. Assuming a conservative scenario where WASP-12 is observed in the short observation mode for 2 months, this will amount to $\sim$50 orbits of the planet leading to a precision of $\sim$30\,ppm/2min. At this noise level in our optical simulations, we recovered a 4.3$\sigma$ detection with \hf{}$=1.59^{+0.40}_{-0.37}$ from the phase curve fit and a 3.6$\sigma$ detection with \hf{}$=1.54\pm0.42$ from the transit fit. Fig.\,\ref{fig:plato_pcfit_30ppm} shows the best-fit models to the simulated \plato observations. \plato 2 – 3 yrs long-pointing observations of UHJs susceptible to deformation will provide exquisite photometric precision that will only be limited by stellar noise and limb darkening modeling.

\medskip
\noindent\textbullet \, \textbf{\tess}: We estimate the precision of \tess from its actual observations of WASP-12\,b for which it obtained a precision of $\sim$1800\,ppm/2min \citep{Wong2022-WASP-12}. A total of 78 orbits of the planet was observed (across 4 \tess sectors) which improves the precision to $\sim$200\,ppm/2min. At this noise level in our optical simulations, we obtained a $\sim$2.3$\sigma$ detection of \hf from the phase curve fit and a slightly lower $\sim$2$\sigma$ detection from the transit fit. Similar to Fig.\,\ref{fig:jwst_pcfit_50ppm}, Fig.\,\ref{fig:tess_pcfit_200ppm} shows the best-fit models to the simulated \tess observations with 200\,ppm/2min noise. In order to detect deformation with higher significance, more \tess observations will be required to lower the noise level.

\begin{figure*}
    \centering
    \includegraphics[width=\linewidth]{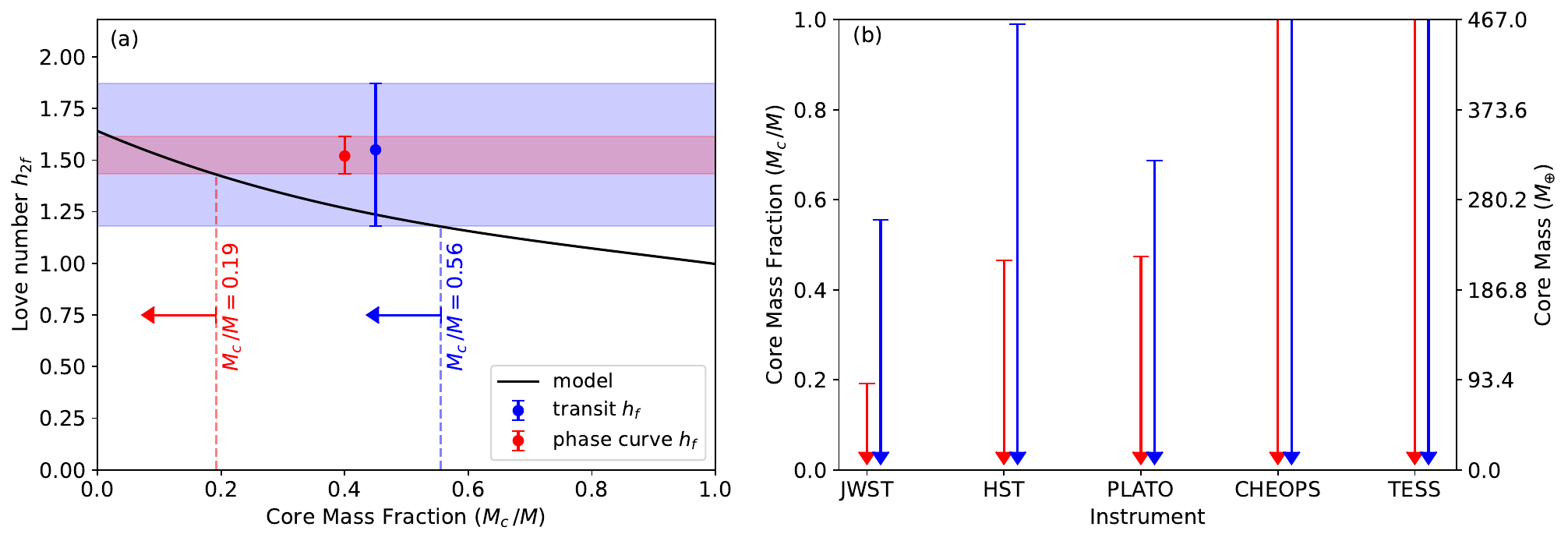}
    \caption{Constraints on the core mass fraction of WASP-12\,b derived from the retrieved \hf. Left: Constraints from \jwst phase curve (red) and transit (blue). The black curve shows the theoretical Love number as a function of core mass fraction for a two-layer model using the formulations of \citet{Buhler2016} and \citet{Helled2013GiantStructur}. Right: Upper limits on core mass fraction based on the precision of retrieved \hf from different instruments. The number of phase curve and transits vists used are: \jwst–1 visit, \hst–3 visits, \plato–50 visits, \cheops–28 visits, \tess–78 visits.}
    \label{fig:core_mass_fraction}
\end{figure*}

\medskip
\noindent\textbullet \, \textbf{\cheops}: Using the \cheops Exposure Time Calculator\footnote{\href{https://cheops.unige.ch/pht2/exposure-time-calculator/}{https://cheops.unige.ch/pht2/exposure-time-calculator/}}, we estimated a photometric precision of 450\,ppm/2min\footnote{consistent with actual \cheops observation of WASP-12\,b \citep{akinsanmi23_prep}} for observations of WASP-12\,b. For a month-long observation of WASP-12\,b (or 28 accumulated phase curves), a precision of 85\,ppm/2min can be attained. At this noise level in our optical simulations, we recovered a 3$\sigma$ detection of \hf from the phase curve fit and 2.5$\sigma$ from the transit fit. Fig.\,\ref{fig:cheops_pcfit_85ppm} shows the best-fit models to the simulated \cheops observations.

\medskip
\noindent\textbullet \, \textbf{HST}: Observations of WASP-12 have been observed with different \hst instruments. Using the G141 grism of \hst/\wfc, \citep{Kreidberg2015AComposition} obtained a precision equivalent to $\sim$110\,ppm/2min on 3 transit visits. Phase curve and transit simulations at this noise level result in a 4.3$\sigma$ and 2.6$\sigma$ detection of \hf, respectively. We see that a noise level of 110\,ppm for \wfc constrains \hf better than 85\,ppm from \cheops due to the larger NIR phase curve amplitude (Fig.\,\ref{fig:hst_pcfit_110ppm}). Doubling the number of visits to 6 improves the detection significance in the phase curve and transit to 5.6$\sigma$ and 3$\sigma$, respectively.

\medskip
\noindent We note gaps within the data (particularly for \hst and \cheops) will reduce the deformation detection significance estimated for the different instruments or increase the number of visits required to reach a certain detection significance. For example, even after combining 3 \hst/\wfc-G141 transit visits of WASP-12\,b in \citet{Kreidberg2015AComposition}, the ingress and egress phases still have poor coverage, making it challenging to measure tidal deformation from these datasets within a reasonable number of \hst visits. However, observations can be scheduled so that the combined visits attain better phase coverage. Instrumental and/or astrophysical noise (e.g., spacecraft jitter or stellar activity) can also impact the detection of deformation from the light curves. These will have to be mitigated while analyzing the data using methods such as polynomial decorrelations, Gaussian Processes \citep{Barros}, or wavelets \citep{csizmadia_wavelets2023}.

\section{Constraining the core mass}
\label{sect:core_mass}
As mentioned previously, measuring the Love number of a planet sheds insight into its interior structure. For instance, \citet{Baumeister2020} found that adding the Love number as input to interior structure models significantly decreases the degeneracy of the possible interior structure. Furthermore, simulations of the interior structure of low mass planets by \citep{Baumeister2023ExoMDN:Networks} found that a Love number precision of 10$\sigma$ can constrain the core and mantle size of an Earth analog to $\sim$13\% of its true value. Here we show how the retrieved value of \hf from the transit and phase curve fits of the different instruments can be used to place constraints on the core mass fraction $M_c\,/M$ of the planet. To do this, we calculate the theoretical value of \hf as a function of $M_c\,/M$ using Eqs.\,1 and 2 of \cite{Buhler2016} and adopting the radial density profile in Eq.\,7 of \citet{Helled2013GiantStructur} which consists of a constant density core and a stratified density envelope. 

Figure\,\ref{fig:core_mass_fraction}a shows the dependence of \hf on $M_c\,/M$. As $M_c\,/M$ increases, the planet is more centrally condensed and less homogeneous which causes \hf to decrease. We overplot the retrieved values of \hf from the transit and phase curve fits of the \jwst \nirspec-PRISM simulations (\S\,\ref{sect:compare_dect}). We see from the $1\sigma$ lower limits of \hf that the phase curve fit constrains the maximum core mass fraction of the planet as 0.19 whereas the larger \hf uncertainties from the transit fit give a maximum $M_c\,/M$ of 0.56. Figure\,\ref{fig:core_mass_fraction}b compares the maximum core mass fraction derived from the retrieved \hf from the phase curve and transit fit of the different instruments. We see that the \jwst phase curve fit provides the tightest constraint on the core mass of the planet followed by the \hst and \plato phase curve fits, and then the \jwst and \plato transit fits. We see that the <$3\sigma$ \hf measurements from \cheops and \tess provides no constraint on the core mass with a maximum core mass being equal to the planetary mass. We note that although the measurements from the different instruments are still consistent with a zero core mass, more precise \hf measurements (e.g. from more than 1 \jwst phase curve or \plato long-pointing) will allow to also determine the lower limit of $M_c\,/M$ definitively constraining the planetary core and improving interior structure models.

\section{Bias on derived system parameters}
\label{sect:par_bias}
\begin{figure*}
    \centering
    \includegraphics[width=0.8\linewidth]{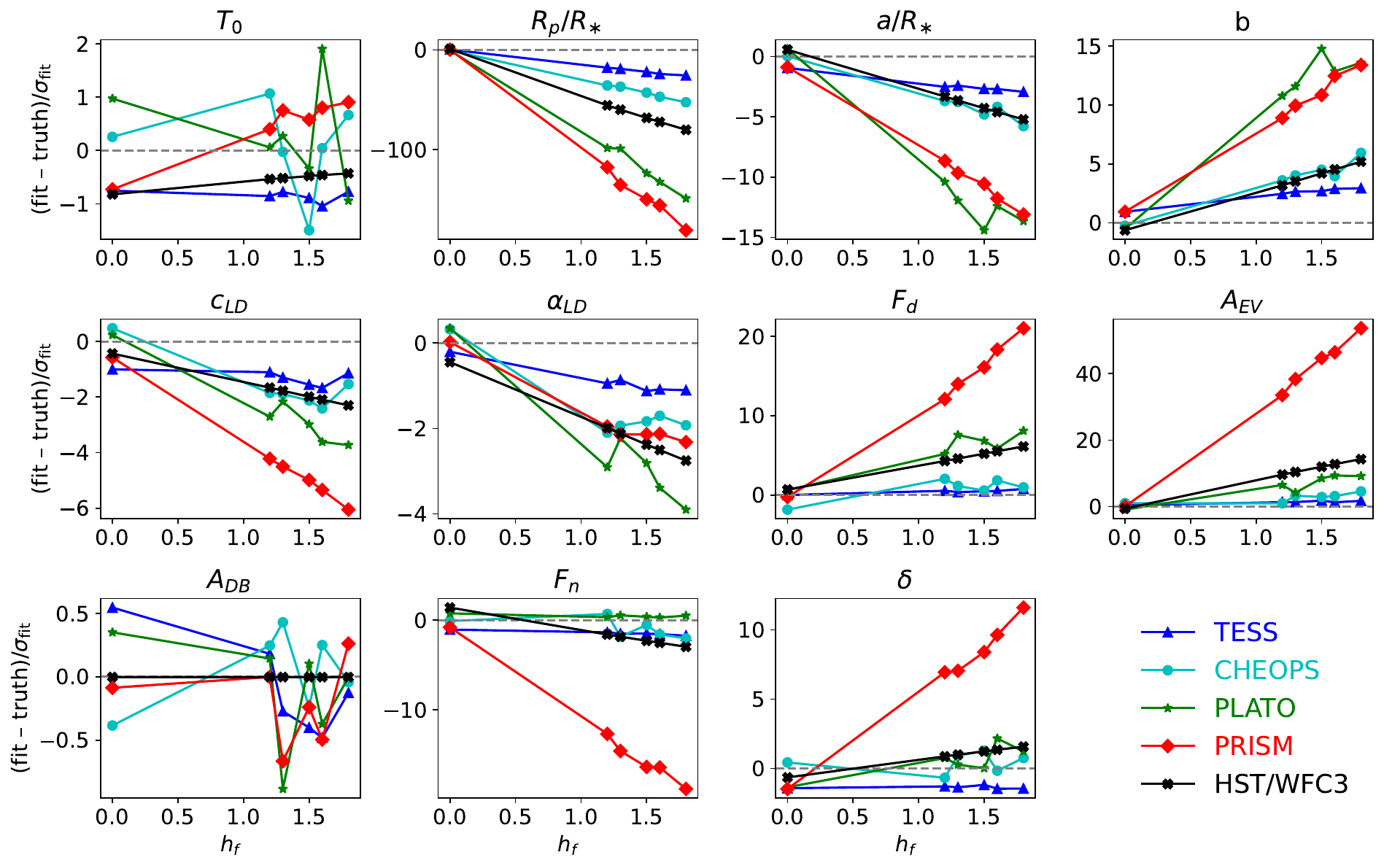}
    \caption{Parameter biases from fitting a spherical planet model to simulated phase curve of deformed planets of different \hf. The parameter names are given in Table\,\ref{tab:sim_wasp12}. For each parameter, we compare the deviation of the fitted spherical planet parameters (fit) from the true input parameters of the deformed planet (truth) and scale this by the standard deviation of the fit ($\sigma_{\mathrm{fit}}$). The colors show the biases for the different instruments.}
    \label{fig:biases}
\end{figure*}
Parameters of the model adopted in fitting observations are often correlated with one another, making it possible for other phase curve parameters to dampen or entirely mask the deformation signal depending on the noise level of the light curve. Assuming sphericity when fitting the phase curve of a deformed planet could introduce biases in the estimation of various other parameters. The deformation then manifests itself as an astrophysical source of systematic error in the measurement of the parameters. 

We investigate the parameter biases by fitting a spherical planet model to the simulated \tess, \plato, \cheops, \hst, and PRISM deformed planet phase curves with increasing values of \hf up to 1.8. The noise level of each instrument estimated in Section\,\ref{sect:compare_dect} was added to the simulated light curves. For each parameter, we compare the obtained posterior to the true simulated value by calculating the number of standard deviations by which the median posterior value differs from the true value i.e. (fit – truth)/$\sigma_{\mathrm{fit}}$. This normalization allows us to easily compare the parameter deviation for each instrument. 

The result of the fits is shown in Fig.\,\ref{fig:biases}. As expected, for \hf=0 (spherical planet) the fitted parameters are close to (<1$\sigma$) the true simulated values leading to deviations close to zero. A zero deviation will also be obtained for cases where the deformed planet phase curve is indistinguishable from that of a spherical planet, due to high noise level in the light curve compared to the deformation signal. 

For all instruments, we observe significant biases in the estimation of $R_p/R_{\star}$, $a/R_{\star}$, and $b$, which all show clear trends with \hf. Since a tidally deformed planet projects only a small cross-section of its shape during transit, the best-fit spherical planet model has a smaller $R_p$ than the volumetric radius \Rv of the ellipsoid. Therefore, as \hf increases and the planet is more deformed, it projects an even smaller cross-sectional area, which requires a smaller spherical $R_p$ to match the transit depth. We find that the derived $R_p/R_{\star}$ increasingly deviates from (underestimates) the simulated $R_\mathrm{v}/R_{\star}$ as a function of \hf. At the highest simulated \hf value of 1.8, $R_p/R_{\star}$ deviates by up to $25\sigma$ for the \tess fit, $50\sigma$ for the \cheops fit, and $>$80$\sigma$ for the \hst, \plato and PRISM phase curve fits which have the most precise radius measurements. When combined with the mass estimate from radial velocity (RV) measurements, the radius deviation can thus lead to an overestimation of the density of the planet by more than 10\% if sphericity is assumed \citep{burton,correia14, Barros2022}. As the precision of RV mass estimates increase, the bias in derived densities due to tidal deformation will constitute a bottleneck in constraining the composition of exoplanets susceptible to deformation \citep{Berardo2022TidalCompositions}.

Both $a/R_{\star}$ and $b$ deviate at \hf{}=1.8 by $>$2.5$\sigma$ for \tess, \cheops, and \hst, while the higher precision phase curves of \plato and PRISM deviate by $>$12$\sigma$. These parameters deviate from the true inputs in order to compensate for the longer transit duration of the deformed planet \citep{Arkhypov2021}. The limb darkening parameters $c_{\mathrm{LD}}$ and $\alpha_{\mathrm{LD}}$ also show biases as these parameters attempt to absorb the deformation signal concentrated around ingress and egress. 

Biases are also observed in the dayside and nightside fluxes, $F_d$ and $F_n$, which show opposing trends as a function of \hf. Both parameters act together to increase the amplitude $A_{\mathrm{atm}}$ of the atmospheric phase variation in order to account for the additional flux from the deformed planet at phases with larger projected areas compared to the sphere (see Fig.\,\ref{fig:jwst_pcfit_50ppm}). Biases on these parameters affect the temperatures derived for the hemispheres of the planet. The temperature of the dayside is overestimated and the nightside is underestimated, which results in a larger day-to-night contrast. The deviations in $F_d$ and $F_n$ are up to 20$\sigma$ at \hf{}=1.8 for the PRISM fit where the amplitudes of these parameters are much higher than the light curve scatter. Converting the dayside and nightside fluxes to temperatures\footnote{Using a BT-Settl stellar model \citep{Allard2012ModelsExoplanets} for WASP-12 and a blackbody for the planet} ($T_d$ and $T_n$), we find that their deviations lead to a 5.4$\sigma$ overestimation of $T_d$ and a 13$\sigma$ underestimation of $T_n$ for PRISM, and $\lesssim$3$\sigma$ deviations for \hst. For the optical phase curves with lower simulated values of $F_d$\,=\,466\,ppm and $F_n$\,=\,0\,ppm, deviations in the derived $F_d$ decrease to $\sim$8$\sigma$ ($T_d$ deviation of $2.2\sigma$) for \plato and $<$2$\sigma$ for \tess and \cheops ($T_d$ deviations of <$1\sigma$) whereas deviations in $F_n$ are close to zero.

The hotspot offset $\delta$ also exhibits a positive deviation of up to 11$\sigma$ at \hf{}=1.8 for the PRISM phase curves with the largest simulated offset of 20$\degree$. The larger eastward hotspot of the spherical planet attempts to compensate for the deformation-induced flux before eclipse. However, the optical simulations with small offsets of 6.2$\degree$ showed only $<$1.5$\sigma$ deviations. As mentioned in section\,\ref{sect:compare_dect}, the amplitude of ellipsoidal variation amplitude increases in an attempt to compensate for the deformation contribution. This leads to a positive deviation from the true value by up to $50\sigma$ at \hf{}=1.8 for the PRISM fit, $\sim$10$\sigma$ for the \hst and \plato fits, $5\sigma$ for the \cheops fit, and $2\sigma$ for the \tess fit. There are no significant deviations in the derived transit time and Doppler beaming amplitude.

In general, the PRISM phase curve shows the largest parameter deviations due to the high precision of \jwst data and the large amplitude deformation signal that the spherical planet model parameters try to compensate for. Conversely, the \tess phase curve shows the smallest parameter deviations due to the high noise level of the simulated observations, which makes the phase curve similar to that of a spherical planet and makes detecting deformation challenging. Nevertheless,  we see that even for such low or non-detection cases, deformation can still lead to significant deviations (ranging from 3 – 25$\sigma$) in some of the derived parameters. Therefore, we recommend accounting for planetary deformation when fitting the phase curves of UHJs by multiplying the planet's atmospheric phase variation by the phase-dependent projected area of the ellipsoid (Eq.\,\ref{eqn:varying_area}). This introduces a single new parameter \hf to fit, if the mass ratio is kept fixed to RV values, which allows marginalizing over the possible shapes of the planet. In analyzing the \texttt{HST} and \spitzer phase curves of WASP-103\,b, \citet{Kreidberg2018a} accounted for the planet's deformation by similarly including the phase-dependent projected ellipsoid area in their model. However, instead of fitting for shape parameters of the ellipsoid, the planet shape was fixed using tabulated predictions in \citet{Leconte2011} assuming a planetary radius and age.

\section{Conclusions}
\label{sect:conclusions}
In this work, we have investigated the effects of tidal deformation on the phase curves of UHJs orbiting close to the Roche limit of their stars. We expand on previous works \citep[e.g.,][]{Leconte2011,burton,correia14,akin19,Hellard2019} studying tidal deformation in transit observations by introducing a simple approach to additionally account for tidal deformation when modeling phase curve observations of UHJs. Similar to \citet{correia14} and \citet{akin19}, our model describes the planet as a triaxial ellipsoid parameterized by the Love number \hf, thereby allowing us to estimate its value from fitting observations. In addition to the signature of deformation in transit caused by the varying projected shape and area of the planet, we showed, using numerical and analytical calculations, that the planet's deformation also manifests itself outside of transit for the same reason. Therefore, the extended phase coverage of the deformation signal in phase curves allows to better probe the planet's shape.

Using injection–retrieval simulations, we find that detecting tidal deformation is more favorable at NIR passbands compared to the optical as we are able to measure \hf significantly at higher noise levels in the NIR. Furthermore, NIR phase curve observations significantly improve the deformation detection compared to transit-only observations due to larger planetary phase curve amplitudes in the NIR. For the precision expected from several \jwst instruments and modes, we find that the measurement of \hf for WASP-12\,b improves from a precision of $\sim$4$\sigma$ from one transit-only analysis to 17$\sigma$ from a single phase curve. Combining several transits equivalent to the phase curve time investment still does not attain the \hf measurement significance of a phase curve.

The high \hf precision obtainable from \jwst phase curves of relevant targets will provide unprecedented constraints on the planet's interior structure (e.g., core mass fraction \citealt{Buhler2016, kramm12}). Additionally, spectrophotometric light curves from \hst and \jwst will allow us to verify that the derived \hf is not wavelength dependent, effectively disentangling tidal deformation from atmospheric phenomena that are chromatic.

Our results also show that detecting deformation from phase curves is relatively unaffected by the uncertainty in the limb darkening profile, as long as there is a high-amplitude deformation signature outside the transit phases, as obtainable for NIR observations. Finally, we showed that the assumption of sphericity when analyzing the phase curve of UHJs can lead to biases in several derived parameters, thereby significantly limiting the ability to properly characterize these targets (their radii, densities, dayside and nightside temperatures, among others). We thus recommend accounting for deformation by modifying the planet's phase variation by the phase-dependent projected area of the ellipsoid, even for cases where the deformation is barely or not detectable. Atmospheric studies of UHJs through transmission spectroscopy could also be affected by tidal deformation. \citet{Lendl2017} suggested that planet deformation could be the cause of the enhanced features measured in the transmission spectra of WASP-103\,b, but this remains to be understood and can be investigated using our ellipsoidal planet model.

\begin{acknowledgements}
This work has been carried out within the framework of the National Centre of Competence in Research PlanetS supported by the Swiss National Science Foundation under grants 51NF40\_182901 and 51NF40\_205606. The ML and BA acknowledge the financial support of the Swiss National Science Foundation under grant number PCEFP2\_194576. S.C.C.B. acknowledges support from FCT -Fundação para a Ciência e a Tecnologia through national funds and by FEDER through COMPETE2020 - Programa Operacional Competividade e Internacionalização by these grants: UIDB/04434/2020; UIDP/04434/2020, 2022.06962.PTDC.
\end{acknowledgements}

\bibliographystyle{aa} 
\bibliography{aanda}

\begin{thebibliography}{67}
\expandafter\ifx\csname natexlab\endcsname\relax\def\natexlab#1{#1}\fi

\bibitem[{Akinsanmi {et~al.}(2019)Akinsanmi, Barros, Santos, Correia, Maxted,
  Bou{\'{e}}, \& Laskar}]{akin19}
Akinsanmi, B., Barros, S.~C., Santos, N.~C., {et~al.} 2019, Astronomy and
  Astrophysics, 621, A117

\bibitem[{{Akinsanmi et al.}(in prep.)}]{akinsanmi23_prep}
{Akinsanmi et al.} in prep.

\bibitem[{Allard {et~al.}(2012)Allard, Homeier, Freytag, Allard, Homeier, \&
  Freytag}]{Allard2012ModelsExoplanets}
Allard, F., Homeier, D., Freytag, B., {et~al.} 2012, RSPTA, 370, 2765

\bibitem[{Arkhypov {et~al.}(2021)Arkhypov, Khodachenko, \&
  Hanslmeier}]{Arkhypov2021}
Arkhypov, O.~V., Khodachenko, M.~L., \& Hanslmeier, A. 2021, Astronomy and
  Astrophysics, 646, A136

\bibitem[{Barros {et~al.}(2022)Barros, Akinsanmi, Bou{\'{e}}, Smith, Laskar,
  Ulmer-Moll, Lillo-Box, Queloz, Cameron, Sousa, Ehrenreich, Hooton, Bruno,
  Demory, Correia, Demangeon, Wilson, Bonfanti, Hoyer, Alibert, Alonso,
  Escud{\'{e}}, Barbato, B{\'{a}}rczy, Barrado, Baumjohann, Beck, Beck, Benz,
  Bergomi, Billot, Bonfils, Bouchy, Brandeker, Broeg, Cabrera, Cessa, Charnoz,
  Damme, Davies, Deleuil, Deline, Delrez, Erikson, Fortier, Fossati, Fridlund,
  Gandolfi, Mu{\~{n}}oz, Gillon, G{\"{u}}del, Isaak, Heng, Kiss, des Etangs,
  Lendl, Lovis, Magrin, Nascimbeni, Maxted, Olofsson, Ottensamer, Pagano,
  Pall{\'{e}}, Parviainen, Peter, Piotto, Pollacco, Ragazzoni, Rando, Rauer,
  Ribas, Santos, Scandariato, S{\'{e}}gransan, Simon, Steller, Szab{\'{o}},
  Thomas, Udry, Ulmer, Van~Grootel, \& Walton}]{Barros2022}
Barros, S. C.~C., Akinsanmi, B., Bou{\'{e}}, G., {et~al.} 2022, Astronomy {\&}
  Astrophysics, 657, A52

\bibitem[{Barros {et~al.}(2012)Barros, Pollacco, Gibson, Keenan, Skillen, \&
  Steele}]{Barros}
Barros, S. C.~C., Pollacco, D.~L., Gibson, N.~P., {et~al.} 2012, Monthly
  Notices of the Royal Astronomical Society, 419, 1248

\bibitem[{Batygin {et~al.}(2009)Batygin, Bodenheimer, \&
  Laughlin}]{Batygin2009}
Batygin, K., Bodenheimer, P., \& Laughlin, G. 2009, Astrophysical Journal, 704,
  L49

\bibitem[{Baumeister {et~al.}(2020)Baumeister, Padovan, Tosi, Montavon,
  Nettelmann, MacKenzie, \& Godolt}]{Baumeister2020}
Baumeister, P., Padovan, S., Tosi, N., {et~al.} 2020, The Astrophysical
  Journal, 889, 42

\bibitem[{Baumeister \& Tosi(2023)}]{Baumeister2023ExoMDN:Networks}
Baumeister, P. \& Tosi, N. 2023, Astronomy {\&} Astrophysics, 676, A106

\bibitem[{Bell {et~al.}(2019)Bell, Zhang, Cubillos, Dang, Fossati, Todorov,
  Cowan, Deming, Zellem, Stevenson, Crossfield, Dobbs-Dixon, Fortney, Knutson,
  \& Line}]{bell-2019-WASP-12}
Bell, T.~J., Zhang, M., Cubillos, P.~E., {et~al.} 2019, Monthly Notices of the
  Royal Astronomical Society, 489, 1995

\bibitem[{Berardo \& de~Wit(2022)}]{Berardo-Dewitt2022}
Berardo, D. \& de~Wit, J. 2022, The Astrophysical Journal, 935, 178

\bibitem[{Berardo \& De~Wit(2022)}]{Berardo2022TidalCompositions}
Berardo, D. \& De~Wit, J. 2022, The Astrophysical Journal, 941, 155

\bibitem[{Bourrier {et~al.}(2020)Bourrier, Kitzmann, Kuntzer, Nascimbeni,
  Lendl, Lavie, Hoeijmakers, Pino, Ehrenreich, Heng, Allart, Cegla, Dumusque,
  Melo, Astudillo-Defru, Caldwell, Cretignier, Giles, Henze, Jenkins, Lovis,
  Murgas, Pepe, Ricker, Rose, Seager, Segransan, Su{\'{a}}rez-Mascare{\~{n}}o,
  Udry, Vanderspek, \& Wyttenbach}]{Bourrier2020}
Bourrier, V., Kitzmann, D., Kuntzer, T., {et~al.} 2020, Astronomy and
  Astrophysics, 637

\bibitem[{Buhler {et~al.}(2016)Buhler, Knutson, Batygin, Fulton, Fortney,
  Burrows, \& Wong}]{Buhler2016}
Buhler, P.~B., Knutson, H.~A., Batygin, K., {et~al.} 2016, The Astrophysical
  Journal, 821, 26

\bibitem[{Burton {et~al.}(2014)Burton, Watson, Fitzsimmons, Pollacco, Moulds,
  Littlefair, \& Wheatley}]{burton}
Burton, J.~R., Watson, C.~A., Fitzsimmons, A., {et~al.} 2014, The Astrophysical
  Journal, 789, 113

\bibitem[{Claret \& Southworth(2022)}]{Claret_southworth2022_power2_ldcs}
Claret, A. \& Southworth, J. 2022, Astronomy {\&} Astrophysics, 664, A128

\bibitem[{Collins {et~al.}(2017)Collins, Kielkopf, \&
  Stassun}]{Collins2017TRANSITPLANETS}
Collins, K.~A., Kielkopf, J.~F., \& Stassun, K.~G. 2017, The Astronomical
  Journal, 153

\bibitem[{Correia(2014)}]{correia14}
Correia, A. C.~M. 2014, Astronomy {\&} Astrophysics, 570, L5

\bibitem[{Cowan {et~al.}(2012)Cowan, Machalek, Croll, Shekhtman, Burrows,
  Deming, Greene, \& Hora}]{cowan2012-wasp-12}
Cowan, N.~B., Machalek, P., Croll, B., {et~al.} 2012, The Astrophysical
  Journal, 747, 82

\bibitem[{{Csizmadia} {et~al.}(2023){Csizmadia}, {Smith}, {K{\'a}lm{\'a}n},
  {Cabrera}, {Klagyivik}, {Chaushev}, \& {Lam}}]{csizmadia_wavelets2023}
{Csizmadia}, S., {Smith}, A.~M.~S., {K{\'a}lm{\'a}n}, S., {et~al.} 2023, \aap,
  675, A106

\bibitem[{Durante {et~al.}(2020)Durante, Parisi, Serra, Zannoni, Notaro,
  Racioppa, Buccino, Lari, Gomez~Casajus, Iess, Folkner, Tommei, Tortora, \&
  Bolton}]{Durante2020}
Durante, D., Parisi, M., Serra, D., {et~al.} 2020, Geophysical Research
  Letters, 47, e2019GL086572

\bibitem[{Espinoza \& Jones(2021)}]{Espinoza2021ConstrainingSpectroscopy}
Espinoza, N. \& Jones, K. 2021, The Astronomical Journal, 162, 165

\bibitem[{Esteves {et~al.}(2013)Esteves, De~Mooij, \&
  Jayawardhana}]{Esteves2013OPTICALEXOPLANETS}
Esteves, L.~J., De~Mooij, E. J.~W., \& Jayawardhana, R. 2013, The Astrophysical
  Journal, 772, 51

\bibitem[{{Hardy} {et~al.}(2017){Hardy}, {Harrington}, {Hardin}, {Madhusudhan},
  {Loredo}, {Challener}, {Foster}, {Cubillos}, \& {Blecic}}]{hardy_2017}
{Hardy}, R.~A., {Harrington}, J., {Hardin}, M.~R., {et~al.} 2017, \apj, 836,
  143

\bibitem[{Hellard {et~al.}(2019)Hellard, Csizmadia, Padovan, Rauer, Cabrera,
  Sohl, Spohn, \& Breuer}]{Hellard2019}
Hellard, H., Csizmadia, S., Padovan, S., {et~al.} 2019, The Astrophysical
  Journal, 878, 119

\bibitem[{Hellard {et~al.}(2020)Hellard, Csizmadia, Padovan, Sohl, \&
  Rauer}]{Hellard2020}
Hellard, H., Csizmadia, S., Padovan, S., Sohl, F., \& Rauer, H. 2020, The
  Astrophysical Journal, 889, 66

\bibitem[{Helled {et~al.}(2014)Helled, Bodenheimer, Podolak, Boley, Meru,
  Nayakshin, Fortney, Mayer, Alibert, \& Boss}]{Helled2013GiantStructur}
Helled, R., Bodenheimer, P., Podolak, M., {et~al.} 2014, in Protostars and
  Planets VI (University of Arizona Press)

\bibitem[{Helling {et~al.}(2020)Helling, Kawashima, Graham, Samra, Chubb, Min,
  Waters, \& Parmentier}]{Helling2020MineralWASP-43b}
Helling, C., Kawashima, Y., Graham, V., {et~al.} 2020, Astronomy and
  Astrophysics, 641, A178

\bibitem[{Hestroffer(1997)}]{Hestroffer1997}
Hestroffer, D. 1997, Astronomy and Astrophysics, 327, 199

\bibitem[{Hooton {et~al.}(2019)Hooton, de Mooij, Watson, Gibson, Galindo-Guil,
  Clavero, \& Merritt}]{Hooton2019StormsWASP-12b}
Hooton, M.~J., de Mooij, E. J.~W., Watson, C.~A., {et~al.} 2019, Monthly
  Notices of the Royal Astronomical Society, 486, 2397

\bibitem[{Jones \& Espinoza(2020)}]{Jones2020Catwoman:Curves}
Jones, K. \& Espinoza, N. 2020, Journal of Open Source Software, 7, 2382

\bibitem[{Kellermann {et~al.}(2018)Kellermann, Becker, \&
  Redmer}]{Kellermann2018}
Kellermann, C., Becker, A., \& Redmer, R. 2018, Astronomy and Astrophysics,
  615, 39

\bibitem[{Kramm {et~al.}(2012)Kramm, Nettelmann, Fortney, Neuh{\"{a}}user, \&
  Redmer}]{kramm12}
Kramm, U., Nettelmann, N., Fortney, J.~J., Neuh{\"{a}}user, R., \& Redmer, R.
  2012, Astronomy and Astrophysics, 538

\bibitem[{Kramm {et~al.}(2011)Kramm, Nettelmann, Redmer, \&
  Stevenson}]{kramm11}
Kramm, U., Nettelmann, N., Redmer, R., \& Stevenson, D.~J. 2011, Astronomy {\&}
  Astrophysics, 528, A18

\bibitem[{Kreidberg(2018)}]{Kreidberg2018a}
Kreidberg, L. 2018, in Handbook of Exoplanets (Cham: Springer International
  Publishing), 2083--2105

\bibitem[{Kreidberg {et~al.}(2015)Kreidberg, Line, Bean, Stevenson,
  D{\'{e}}sert, Madhusudhan, Fortney, Barstow, Henry, Williamson, \&
  Showman}]{Kreidberg2015AComposition}
Kreidberg, L., Line, M.~R., Bean, J.~L., {et~al.} 2015, The Astrophysical
  Journal, 814, 66

\bibitem[{Kreidberg {et~al.}(2018)Kreidberg, Line, Parmentier, Stevenson,
  Louden, Bonnefoy, Faherty, Henry, Williamson, Stassun, Beatty, Bean, Fortney,
  Showman, D{\'{e}}sert, \& Arcangeli}]{kreid18}
Kreidberg, L., Line, M.~R., Parmentier, V., {et~al.} 2018, {Global climate and
  atmospheric composition of the ultra-hot jupiter wasp-103b from hst and
  spitzer phase curve observations}

\bibitem[{Lainey {et~al.}(2017)Lainey, Jacobson, Tajeddine, Cooper, Murray,
  Robert, Tobie, Guillot, Mathis, Remus, Desmars, Arlot, Cuyper, Dehant, Pascu,
  Thuillot, Poncin-Lafitte, \& Zahn}]{LAINEY2017}
Lainey, V., Jacobson, R.~A., Tajeddine, R., {et~al.} 2017, Icarus, 281, 286

\bibitem[{Leconte {et~al.}(2011{\natexlab{a}})Leconte, Lai, \&
  Chabrier}]{Leconte2011}
Leconte, J., Lai, D., \& Chabrier, G. 2011{\natexlab{a}}, Astronomy {\&}
  Astrophysics, 528, A41

\bibitem[{Leconte {et~al.}(2011{\natexlab{b}})Leconte, Lai, \&
  Chabrier}]{Leconte2011a}
Leconte, J., Lai, D., \& Chabrier, G. 2011{\natexlab{b}}, {Erratum: Distorted,
  non-spherical transiting planets: Impact on the transit depth and on the
  radius determination (Astronomy and Astrophysics (2011) 528 (A41) DOI:
  10.1051/0004-6361/201015811)}

\bibitem[{Lendl {et~al.}(2017)Lendl, Cubillos, Hagelberg, M{\"{u}}ller, Juvan,
  \& Fossati}]{Lendl2017}
Lendl, M., Cubillos, P.~E., Hagelberg, J., {et~al.} 2017, Astronomy and
  Astrophysics, 606

\bibitem[{Loeb \& Gaudi(2003)}]{Loeb2003PeriodicCompanions}
Loeb, A. \& Gaudi, B.~S. 2003, The Astrophysical Journal, 588, L117

\bibitem[{Louden \& Kreidberg(2018)}]{Louden2018SPIDERMAN:Eclipses}
Louden, T. \& Kreidberg, L. 2018, Monthly Notices of the Royal Astronomical
  Society, 477, 2613

\bibitem[{Love(1911)}]{love}
Love, A. E.~H. 1911, Nature, 89, 471

\bibitem[{Maxted(2018)}]{maxted18}
Maxted, P.~F. 2018, Astronomy and Astrophysics, 616, 39

\bibitem[{{Mislis} {et~al.}(2012){Mislis}, {Heller}, {Schmitt}, \&
  {Hodgkin}}]{mislis2012}
{Mislis}, D., {Heller}, R., {Schmitt}, J.~H.~M.~M., \& {Hodgkin}, S. 2012,
  \aap, 538, A4

\bibitem[{Morello {et~al.}(2017)Morello, Tsiaras, Howarth, \&
  Homeier}]{morello}
Morello, G., Tsiaras, A., Howarth, I.~D., \& Homeier, D. 2017, The Astronomical
  Journal, 154, 111

\bibitem[{Morris {et~al.}(1985)Morris, {Morris}, \& L.}]{Morris1985TheStars.}
Morris, S.~L., {Morris}, \& L., S. 1985, ApJ, 295, 143

\bibitem[{Newville {et~al.}(2020)Newville, Otten, Nelson, Ingargiola,
  Stensitzki, Allan, Fox, Carter, {Micha{\l}}, Pustakhod, Ram, {Glenn}, Deil,
  {Stuermer}, Beelen, Frost, Zobrist, {Mark}, Pasquevich, Hansen, Spillane,
  Caldwell, Polloreno, {andrewhannum}, Fraine, {deep-42-thought}, Maier,
  Gamari, Persaud, \& Almarza}]{Newville2020Lmfit/lmfit-py1.0.1}
Newville, M., Otten, R., Nelson, A., {et~al.} 2020

\bibitem[{Padovan {et~al.}(2018)Padovan, Spohn, Baumeister, Tosi, Breuer,
  Csizmadia, Hellard, \& Sohl}]{Padovan2018}
Padovan, S., Spohn, T., Baumeister, P., {et~al.} 2018, Astronomy and
  Astrophysics, 620, A178

\bibitem[{Parviainen(2018)}]{Parviainen2018}
Parviainen, H. 2018, in Handbook of Exoplanets (Cham: Springer International
  Publishing), 1567--1590

\bibitem[{Parviainen {et~al.}(2022)Parviainen, Wilson, Lendl, Kitzmann,
  Pall{\'{e}}, Serrano, Valdes, Benz, Deline, Ehrenreich, Guterman, Heng,
  Demangeon, Bonfanti, Salmon, Singh, Santos, Sousa, Alibert, Alonso, Anglada,
  B{\'{a}}rczy, Navascues, Barros, Baumjohann, Beck, Beck, Billot, Bonfils,
  Brandeker, Broeg, Cabrera, Charnoz, Cameron, Van~Damme, Csizmadia, Davies,
  Deleuil, Delrez, Demory, Erikson, Farinato, Fortier, Fossati, Fridlund,
  Gandolfi, Gillon, G{\"{u}}del, Hoyer, Isaak, Kiss, Kopp, Laskar, Etangs,
  Lovis, Magrin, Maxted, Mecina, Nascimbeni, Olofsson, Ottensamer, Pagano,
  Peter, Piazza, Piotto, Pollacco, Queloz, Ragazzoi, Rando, Rauer, Ribas,
  Scandariato, S{\'{e}}gransan, Simon, Smith, Steller, Szab{\'{o}}, Thomas,
  Udry, VanGrootel, \& Walton}]{parvieinen2022}
Parviainen, H., Wilson, T.~G., Lendl, M., {et~al.} 2022, Astronomy {\&}
  Astrophysics, 668, A93

\bibitem[{Pluriel {et~al.}(2020)Pluriel, Zingales, Leconte, \&
  Parmentier}]{Pluriel2020StrongHeterogeneities}
Pluriel, W., Zingales, T., Leconte, J., \& Parmentier, V. 2020, Astronomy {\&}
  Astrophysics, 636, A66

\bibitem[{Ragozzine \& Wolf(2009)}]{Ragozzine2009}
Ragozzine, D. \& Wolf, A.~S. 2009, The Astrophysical Journal, 698, 1778

\bibitem[{Rauer {et~al.}(2014)Rauer, Catala, Aerts, Appourchaux, Benz,
  Brandeker, Christensen-Dalsgaard, Deleuil, Gizon, Goupil, G{\"{u}}del,
  Janot-Pacheco, Mas-Hesse, Pagano, Piotto, Pollacco, {Santos}, Smith,
  Su{\'{a}}rez, Szab{\'{o}}, Udry, Adibekyan, Alibert, Almenara, Amaro-Seoane,
  Eiff, Asplund, Antonello, Barnes, Baudin, Belkacem, Bergemann, Bihain, Birch,
  Bonfils, Boisse, Bonomo, Borsa, Brand{\~{a}}o, Brocato, Brun, Burleigh,
  Burston, Cabrera, Cassisi, Chaplin, Charpinet, Chiappini, Church, Csizmadia,
  Cunha, Damasso, Davies, Deeg, D{\'{i}}az, Dreizler, Dreyer, Eggenberger,
  Ehrenreich, Eigm{\"{u}}ller, Erikson, Farmer, Feltzing, de~Oliveira~Fialho,
  Figueira, Forveille, Fridlund, Garc{\'{i}}a, Giommi, Giuffrida, Godolt,
  da~Silva, Granzer, Grenfell, Grotsch-Noels, G{\"{u}}nther, Haswell, Hatzes,
  H{\'{e}}brard, Hekker, Helled, Heng, Jenkins, Johansen, Khodachenko,
  Kislyakova, Kley, Kolb, Krivova, Kupka, Lammer, Lanza, Lebreton, Magrin,
  Marcos-Arenal, Marrese, Marques, Martins, Mathis, Mathur, Messina, Miglio,
  Montalban, Montalto, P.~F. G.~Monteiro, Moradi, Moravveji, Mordasini, Morel,
  Mortier, Nascimbeni, Nelson, Nielsen, Noack, Norton, Ofir, Oshagh, Ouazzani,
  P{\'{a}}pics, Parro, Petit, Plez, Poretti, Quirrenbach, Ragazzoni, Raimondo,
  Rainer, Reese, Redmer, Reffert, Rojas-Ayala, Roxburgh, Salmon, Santerne,
  Schneider, Schou, Schuh, Schunker, Silva-Valio, Silvotti, Skillen, Snellen,
  Sohl, Sousa, Sozzetti, Stello, Strassmeier, {\v{S}}vanda, Szab{\'{o}},
  Tkachenko, Valencia, Van~Grootel, Vauclair, Ventura, Wagner, Walton,
  Weingrill, Werner, Wheatley, \& Zwintz}]{rauer}
Rauer, H., Catala, C., Aerts, C., {et~al.} 2014, Experimental Astronomy, 38,
  249

\bibitem[{Rauer \& Heras(2018)}]{Rauer2018}
Rauer, H. \& Heras, A.~M. 2018, in Handbook of Exoplanets (Cham: Springer
  International Publishing), 1309--1330

\bibitem[{{Russell}(1912)}]{russel_1912}
{Russell}, H.~N. 1912, \apj, 35, 315

\bibitem[{Sabadini \& Vermeersen(2004)}]{sabadini}
Sabadini, R. \& Vermeersen, B. 2004, {Global Dynamics of the Earth:
  Applications of Normal Mode Relaxation Theory to Solid-Earth Geophysics}
  (Kluwer Academic Publishers)

\bibitem[{Shporer(2017)}]{Shporer2017-PCreview}
Shporer, A. 2017, Publications of the Astronomical Society of the Pacific, 129,
  072001

\bibitem[{Shporer {et~al.}(2019)Shporer, Wong, Huang, Line, Stassun, Fetherolf,
  Kane, Bouma, Daylan, G{\"{u}}enther, Ricker, Latham, Vanderspek, Seager,
  Winn, Jenkins, Glidden, Berta-Thompson, Ting, Li, \&
  Haworth}]{shporer-2019-WASP18}
Shporer, A., Wong, I., Huang, C.~X., {et~al.} 2019, The Astronomical Journal,
  157, 178

\bibitem[{Speagle(2019)}]{speagle}
Speagle, J.~S. 2019, arXiv e-prints, arXiv:1904.02180

\bibitem[{Stevenson {et~al.}(2014)Stevenson, Bean, Seifahrt, D{\'{e}}sert,
  Madhusudhan, Bergmann, Kreidberg, \& Homeier}]{Stevenson2014TRANSMISSIONm}
Stevenson, K.~B., Bean, J.~L., Seifahrt, A., {et~al.} 2014, The Astronomical
  Journal, 147, 161

\bibitem[{Wahl {et~al.}(2021)Wahl, Thorngren, Lu, \&
  Militzer}]{Wahl2021TidalJupiters}
Wahl, S.~M., Thorngren, D., Lu, T., \& Militzer, B. 2021, The Astrophysical
  Journal, 921, 105

\bibitem[{Wong {et~al.}(2021)Wong, Kitzmann, Shporer, Heng, Fetherolf, Benneke,
  Daylan, Kane, Vanderspek, Seager, Winn, Jenkins, \&
  Ting}]{wong2021_tess2ndyearPCs}
Wong, I., Kitzmann, D., Shporer, A., {et~al.} 2021, The Astronomical Journal,
  162, 127

\bibitem[{Wong {et~al.}(2022)Wong, Shporer, Vissapragada, Greklek-McKeon,
  Knutson, Winn, \& Benneke}]{Wong2022-WASP-12}
Wong, I., Shporer, A., Vissapragada, S., {et~al.} 2022, The Astronomical
  Journal, 163, 175

\bibitem[{Zhang \& Showman(2017)}]{Zhang2017EffectsExoplanets}
Zhang, X. \& Showman, A.~P. 2017, The Astrophysical Journal, 836, 73

\bibitem[{Zieba {et~al.}(2022)Zieba, Zilinskas, Kreidberg, Nguyen, Miguel,
  Cowan, Pierrehumbert, Carone, Dang, Hammond, Louden, Lupu, Malavolta, \&
  Stevenson}]{Zieba2022K2Atmosphere}
Zieba, S., Zilinskas, M., Kreidberg, L., {et~al.} 2022, Astronomy {\&}
  Astrophysics, 664, A79

\end{thebibliography}

\onecolumn

\begin{appendix}

\section{Figures}
 \begin{figure}[ht]
    \centering
    \includegraphics[width=0.6\textwidth]{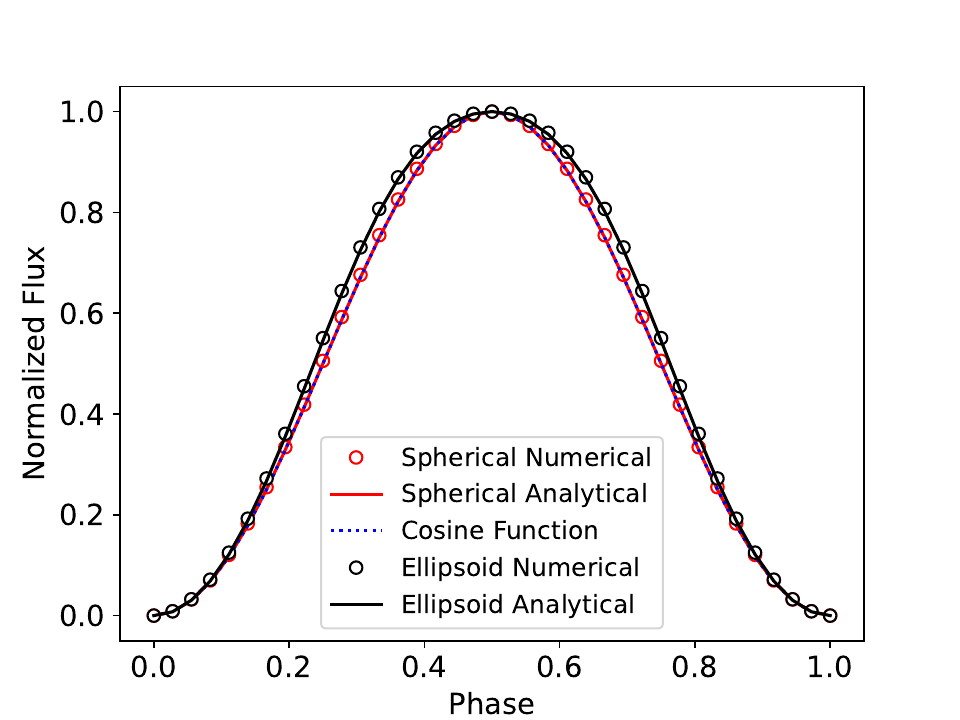}
    \caption{Comparison of the numeric and analytic computation of the planetary phase variation signal. The spherical planet signal is equivalent to a cosine function.}
\label{fig:comp_num_analy}
\end{figure}

 \begin{figure}[ht]
    \centering
    \includegraphics[width=0.6\textwidth]{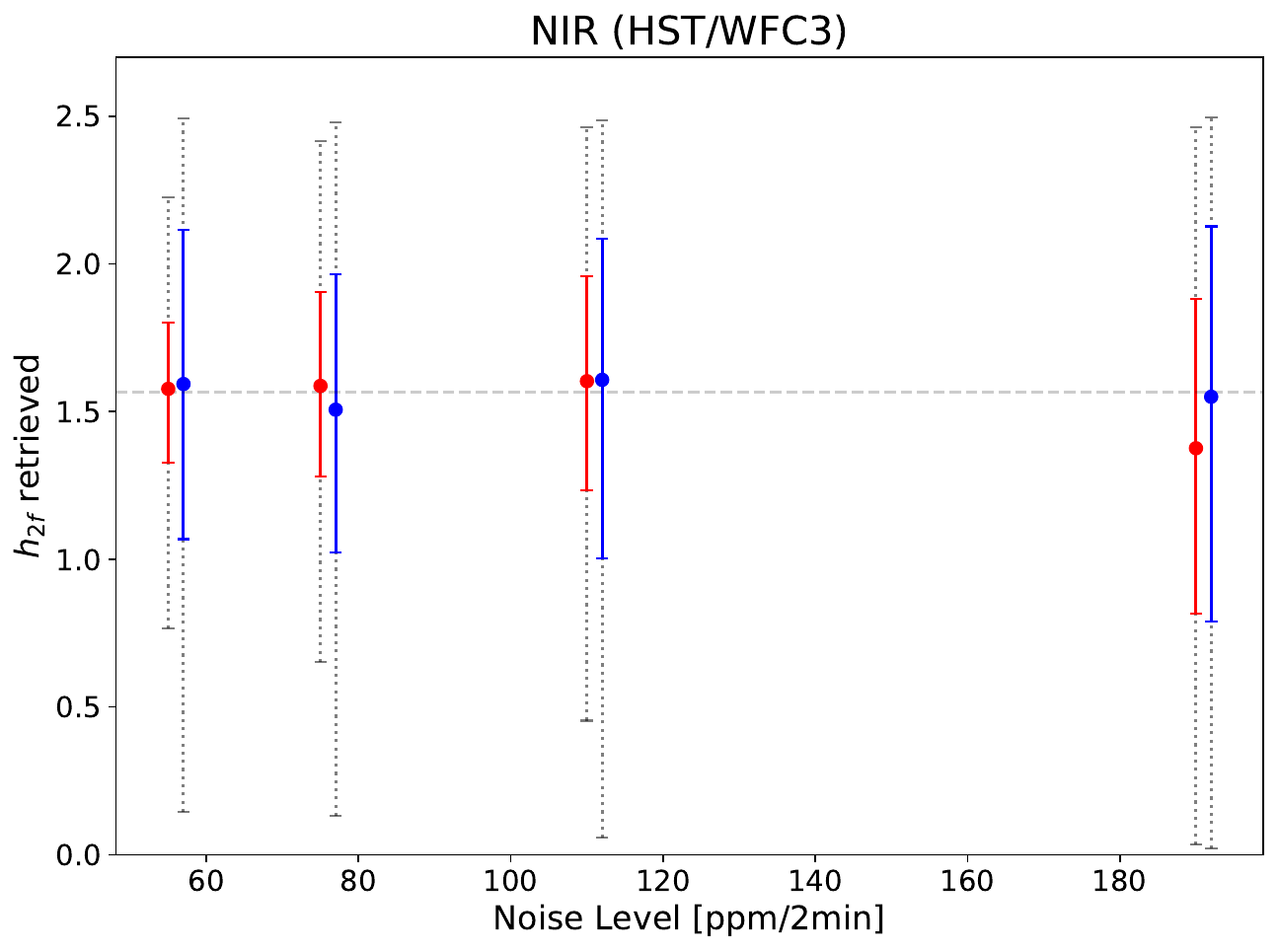}
    \caption{Same as Fig.\,\ref{fig:PCfit} but for \hst/\wfc/ simulations.}
    \label{fig:pc_fit_hst}
\end{figure}

\begin{figure*}[ht]
    \centering
    \includegraphics[width=.8\linewidth]{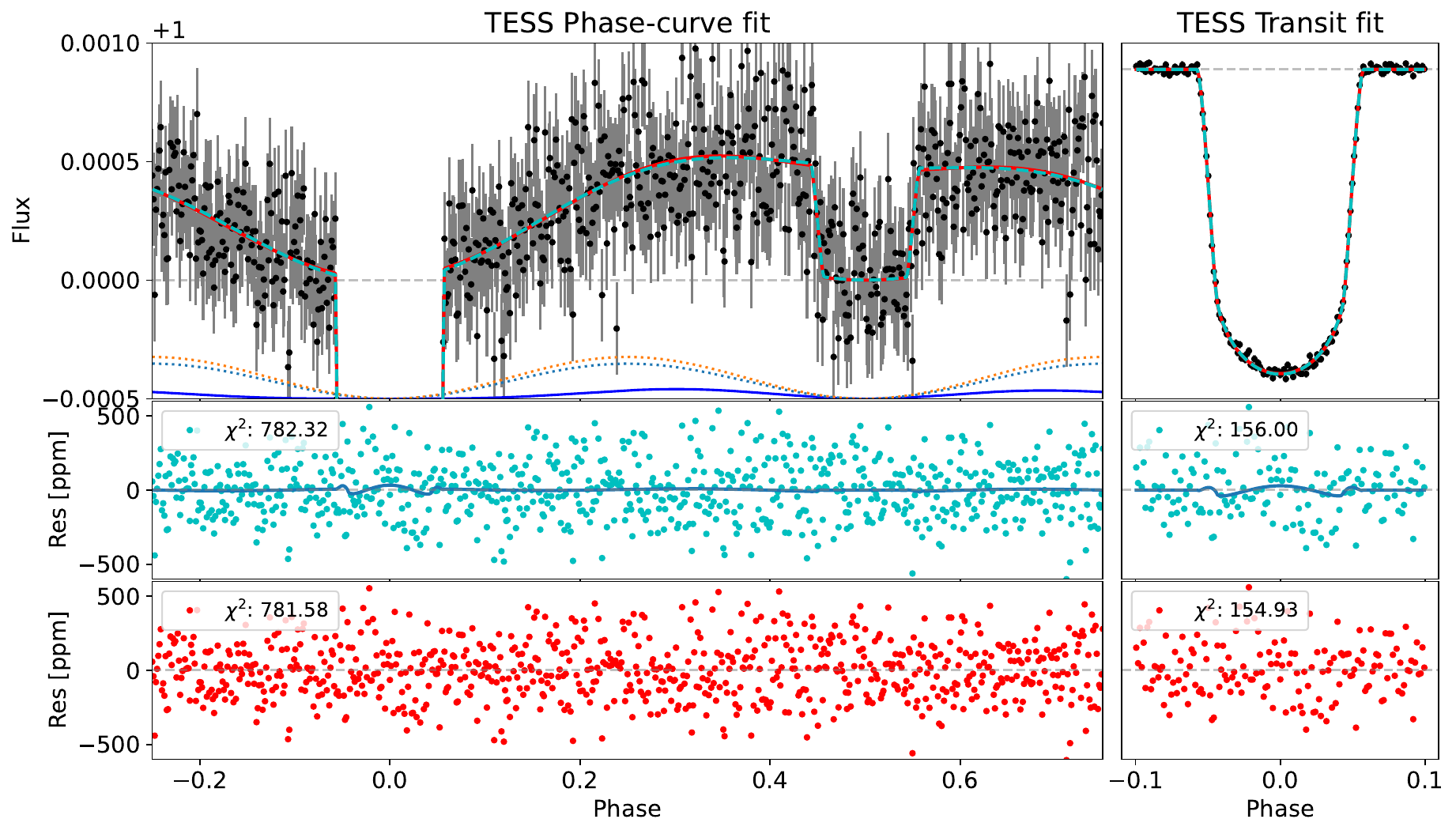}
    \caption{Same as Fig.\,\ref{fig:jwst_pcfit_50ppm} but for simulated 200\,ppm/2min noise observations for \tess.}
    \label{fig:tess_pcfit_200ppm}
\end{figure*}

\begin{figure*}
    \centering
    \includegraphics[width=.8\linewidth]{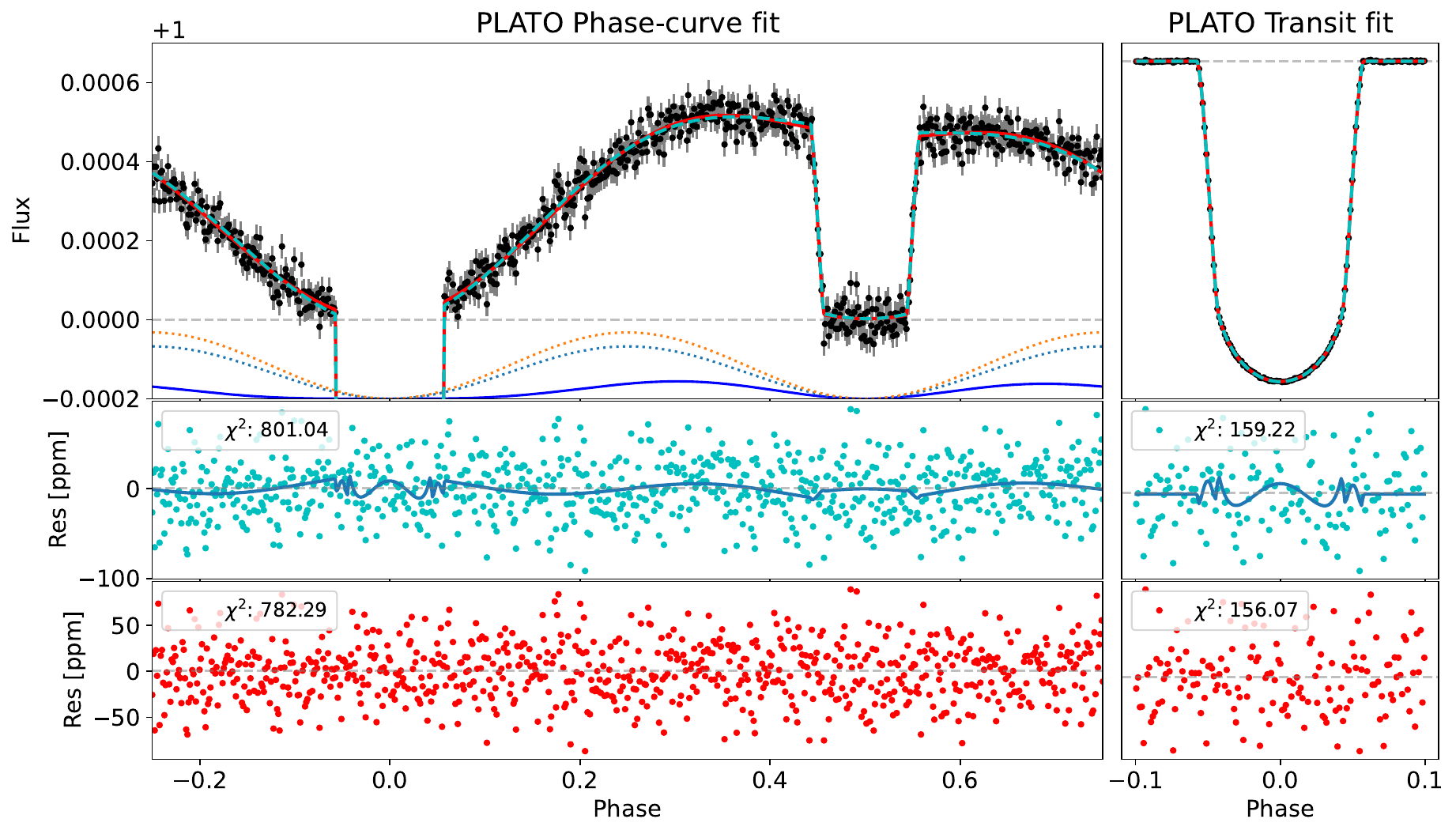}
    \caption{Same as Fig.\,\ref{fig:jwst_pcfit_50ppm} but for simulated 30\,ppm/2min noise observations of \plato.}
    \label{fig:plato_pcfit_30ppm}
\end{figure*}

\begin{figure*}[hb!]
    \centering
    \includegraphics[width=.8\linewidth]{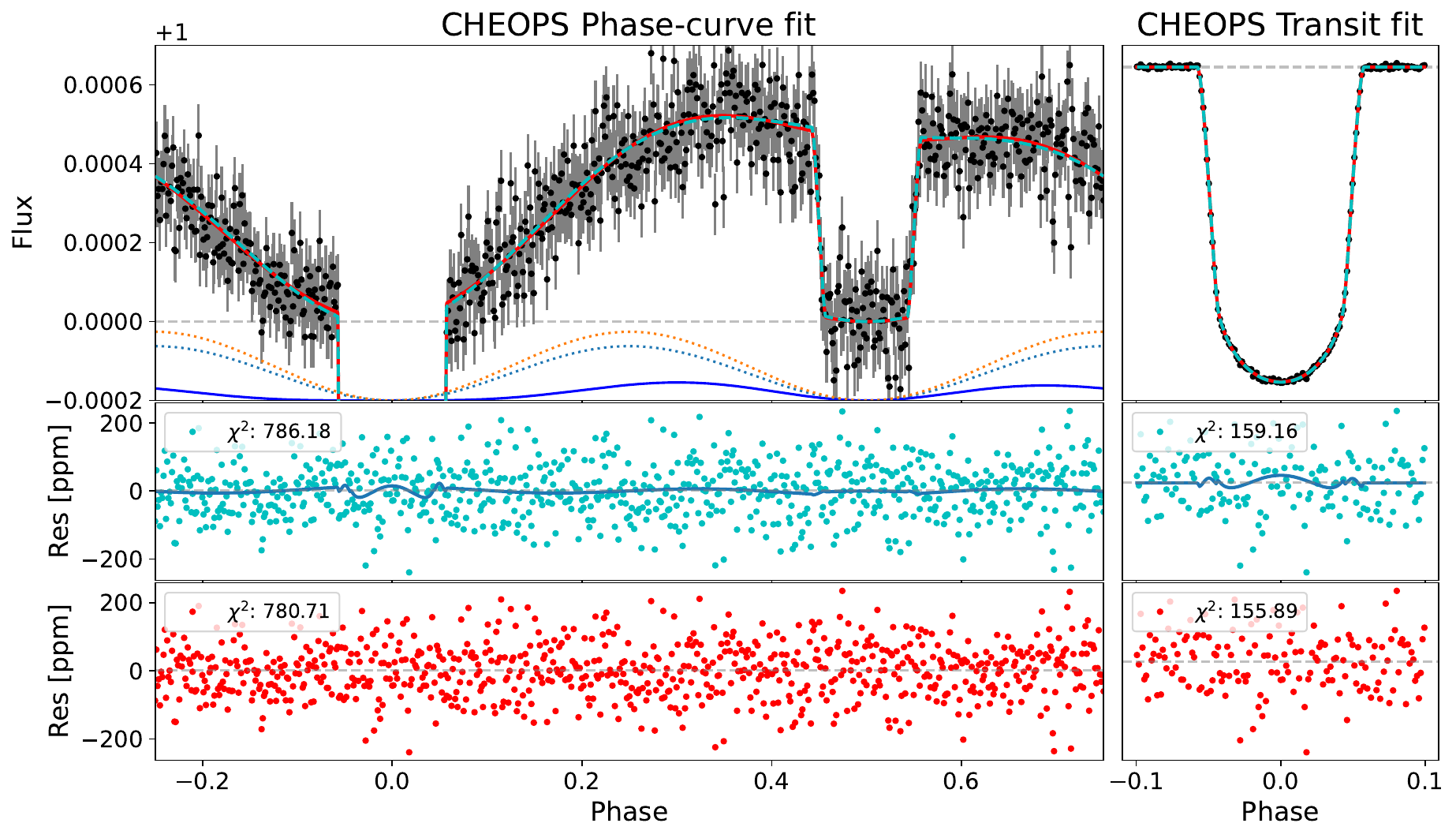}
    \caption{Same as Fig.\,\ref{fig:jwst_pcfit_50ppm} but for simulated 85\,ppm/2min noise observations for \cheops.}
    \label{fig:cheops_pcfit_85ppm}
\end{figure*}

\begin{figure*}[ht]
    \centering
    \includegraphics[width=.8\linewidth]{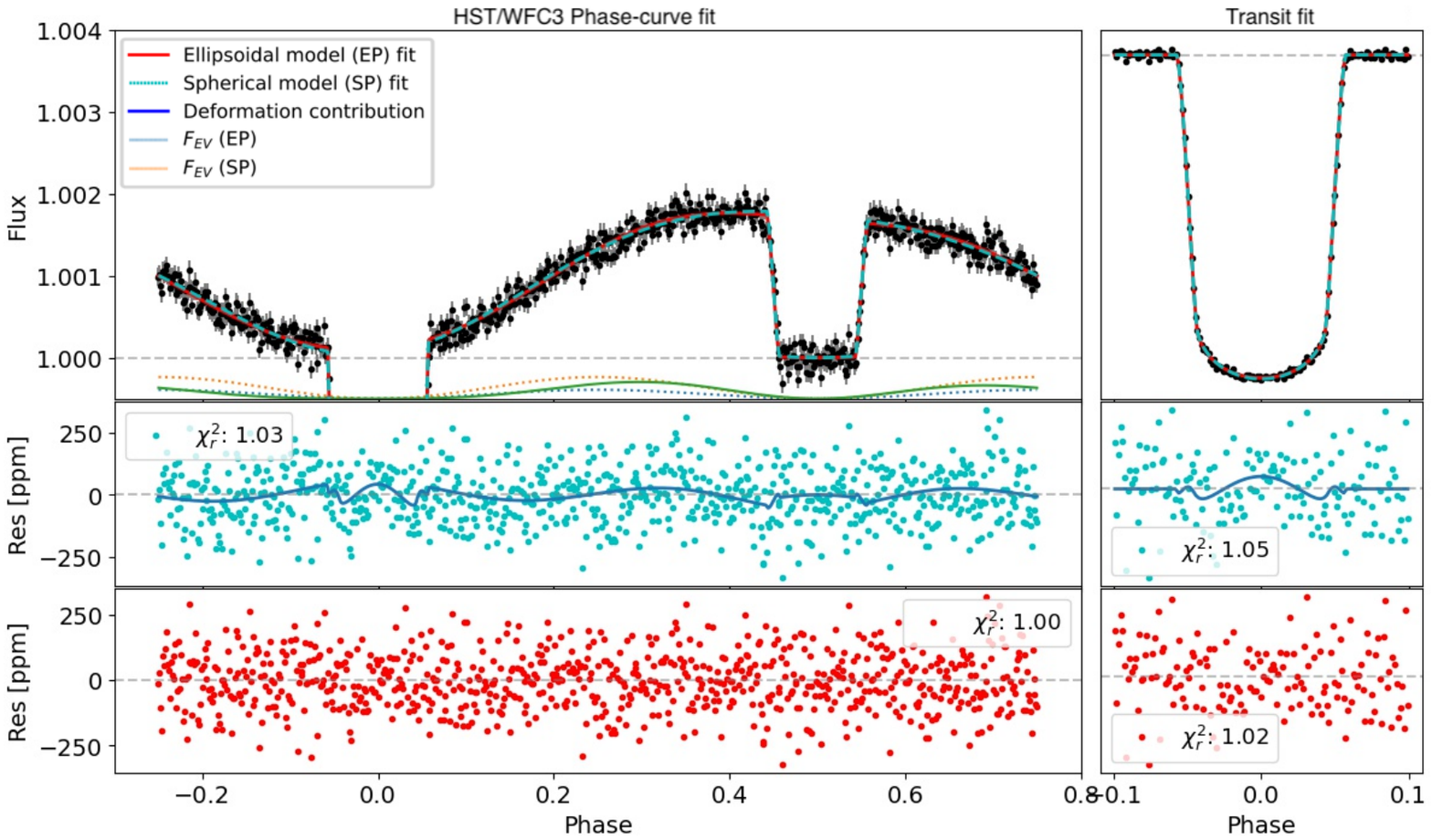}
    \caption{Same as Fig.\,\ref{fig:jwst_pcfit_50ppm} but for simulated 110\,ppm/2min noise observations for \hst/\wfc-G141.}
    \label{fig:hst_pcfit_110ppm}
\end{figure*}

\end{appendix}

\end{document}